\documentclass[11pt,a4paper]{article}
\usepackage{amsmath, amsthm, amssymb} 
\usepackage{amsfonts} 
\usepackage{latexsym}
\usepackage{amsmath,amscd}
\usepackage{graphicx}
\usepackage[margin=1in]{geometry}
\usepackage{makeidx}
\usepackage[english]{babel}
\usepackage{changepage}  
\usepackage{color}
\usepackage{pict2e}
\usepackage{hyperref}
\usepackage{cite}

\newcommand*\xbar[1]{%
  \hbox{%
    \vbox{%
      \hrule height 0.5pt 
      \kern0.5ex
      \hbox{%
        \kern-0.1em
        \ensuremath{#1}%
        \kern-0.1em
      }%
    }%
  }%
}

\def\hybrid{\topmargin -10pt    \oddsidemargin 0pt
        \headheight 0pt \headsep 0pt
       \textwidth 6.25in       
      \textheight 9.5in       
        \marginparwidth .875in
        \parskip 5pt plus 1pt   \jot = 1.5ex}
\hybrid
\theoremstyle{remark}

\theoremstyle{definition}

\newcommand{\be}{\begin{equation}}
\newcommand{\ee}{\end{equation}}
\newcommand{\bea}{\begin{eqnarray}}
\newcommand{\eea}{\end{eqnarray}}

\newcommand{\al}{\alpha}
\renewcommand{\d}{\delta}
\newcommand{\e}{\epsilon}

\newcommand{\g}{\gamma}

\newcommand{\la}{\lambda}

\newcommand{\Om}{\Omega}

\newcommand{\s}{\sigma}

\newcommand{\hlf}{\frac{1}{2}}

\newcommand{\non}{\nonumber}

\newcommand{\p}{\partial}

\newcommand{\rr}{\rightarrow}

\newcommand{\Z}{\mathbb{Z}}
\newcommand{\bz}{\bar{z}}

\newcommand{\SU}{\operatorname{SU}}

\newcommand{\U}{\operatorname{U}}

\newcommand{\lp}{\left(}
\newcommand{\rp}{\right)}
\newcommand{\ls}{\left[}
\newcommand{\rs}{\right]}

\newcommand{\mcI}{\mathcal{I}}
\newcommand{\mcO}{\mathcal{O}}

\newcommand{\wta}{\widetilde{a}}

\newcommand{\wtq}{\widetilde{q}}
\newcommand{\wtx}{\widetilde{x}}

\newcommand{\ov}[1]{{\overline{#1}}}

\begin{document}

\thispagestyle{empty}

\rightline{\small MI-TH-1749}

\vskip 3cm
\noindent
{\LARGE \bf Conformal interfaces between free boson orbifold theories}\\
\vskip .8cm
\begin{center}
\linethickness{.06cm}
\line(1,0){447}
\end{center}
\vskip .8cm
\noindent
{\large \bf Melanie Becker$^\dagger$, Yaniel Cabrera$^\dagger$, Daniel Robbins$^\ddagger$}

\vskip 0.2cm

\begin{tabular}{p{3.5in}p{3.5in}}
\hskip -0.15cm ${}^\dagger${\em    George and Cynthia Mitchell Institute} & \hskip -0.15cm ${}^\ddagger${\em Department of Physics} \\
{\em for Fundamental Physics and Astronomy} & {\em University at Albany} \\
{\em Texas A \&M University} & {\em 1400 Washington Ave.} \\
{\em College Station, TX 77843--4242, USA} & {\em Albany, NY 12222, USA}
\end{tabular}

\vskip 0.5cm
{\tt \hskip -.5cm mbecker, cabrera AT physics.tamu.edu, dgrobbins AT albany.edu}

\vskip 1cm

\vskip0.6cm

\noindent
{\sc Abstract:} We construct a large class of conformal interfaces between two-dimensional $c=1$ conformal field theories describing compact free bosons and their $\Z_2$ orbifolds.  The interfaces are obtained by constructing boundary states in the corresponding $c=2$ product theories and applying the unfolding procedure.  We compute the fusion products for all of these defects, and identify the invertible topological interfaces associated to global symmetries, the interfaces corresponding to marginal deformations, and the interfaces which map the untwisted sector of an orbifold to the invariant states of the parent theory.

\pagebreak

\newpage

\tableofcontents

\section{Introduction}\label{intro}

The consideration of 2D conformal field theories on surfaces with boundaries dates back to the seminal works by John Cardy on the upper half-plane and the strip~\cite{cardy}.  Let's take a theory on the upper half-plane.  The presence of the boundary breaks half of the conformal symmetry of the bulk theory, leaving us with a single copy of the Virasoro algebra.  Along the real axis, a boundary condition must be chosen, and if we want to preserve any of the residual conformal symmetry, we should choose a conformal boundary condition, which means that the holomorphic and antiholomorphic stress tensors $T(z)$ and $\widetilde{T}(\bz)$ agree at the boundary $z=\bz$ when we impose the boundary condition.  If we realize the boundary condition as a boundary state, $|B\rangle\rangle$, then conformality amounts to
\be
\label{eq:ConfBoundState}
\lp L_n-\widetilde{L}_{-n}\rp|B\rangle\rangle=0,\qquad\forall n\in\Z.
\ee
There are in fact more conditions than this that must be satisfied.  The Cardy condition~\cite{Cardy:1989ir} is essentially the expression of modular invariance for the annulus, and is only satisfied by certain linear combinations of states satisfying (\ref{eq:ConfBoundState}).  Further constraints can also be formulated~\cite{Cardy:1991tv,Lewellen:1991tb}, but they are often more difficult to use directly.

The combined bulk and boundary system is called a boundary CFT, and includes new local operators which are restricted to the boundary.  The study of boundary CFTs has been an important field of research in 2D theories with wide applications to D-branes in string theory \cite{Recknagel:1998ih, Brunner:1999, Gaberdiel:2001zq, Gaberdiel:2008rk}.

A generalization of the idea of a conformal boundary is the notion of a conformal interface~\cite{bachas02,bachas07,fuchs07,gaiotto12,Quella:2002CT, brunner03, brunner07, Konechny:2015qla, Graham:2003nc, Fuchs:2015ska}.  In this case we have two CFTs, $\text{CFT}_1$ in the upper half-plane and $\text{CFT}_2$ in the lower half-plane, joined by an interface or domain wall.  We will denote this configuration by $\text{CFT}_1|\text{CFT}_2$.  Again we must specify conditions describing the effect of bringing bulk operators close to the interface, and if we want to preserve the residual conformal symmetry, we must have a particular behavior for the stress tensors, namely that when $T_1(z)-\widetilde{T}_1(\bz)$ approaches the interface from above, it must agree with $T_2(z)-\widetilde{T}_2(\bz)$ approaching from below~\cite{bachas02}.  Analogously to the boundary state $|B\rangle\rangle$, we can formulate the interface as an operator $\mcO$ from the Hilbert space of $\text{CFT}_2$ to the Hilbert space of $\text{CFT}_1$, in which case the condition can be stated as
\be
\lp L^{(1)}_n-\widetilde{L}^{(1)}_{-n}\rp\mcO=\mcO\lp L^{(2)}_n-\widetilde{L}^{(2)}_{-n}\rp,\qquad\forall n\in\Z.
\ee
Two special classes of conformal interfaces solve this condition in particular ways.  Factorized, or totally reflective, interfaces act as a boundary condition on each CFT separately, while topological, or totally transmissive, interfaces preserve the holomorphic and antiholomorphic stress tensors individually.

Conformal boundaries and conformal interfaces are closely intertwined topics.  As with conformal boundaries, conformal interfaces can host local operators restricted to the interface, and interfaces must satisfy various consistency conditions analogous to those for boundaries.  Even more directly, a conformal boundary is a special case of a conformal interface, where we take $\text{CFT}_2$ to be the trivial, or empty, conformal field theory.  Conversely, the folding trick~\cite{affleck} allows us to associate two theories joined by a conformal interface along the real line, $\text{CFT}_1|\text{CFT}_2$, to a tensor product theory $\text{CFT}_1\otimes\overline{\text{CFT}}_2$ defined on the upper half-plane with a conformal boundary along the real line.  The bar indicates that we interchange left- and right-movers in $\text{CFT}_2$~\cite{affleck, bachas07}.  The consistency conditions that must be satisfied by the conformal interface translate to the conformal boundary constraints (Cardy condition and sewing conditions) discussed above for the boundary in the product theory.  The folding trick also applies to quantum field theories which are not necessarily CFTs such as general Landau-Ginzburg models~\cite{brunner07}.

The set of conformal interfaces comes equipped with a product structure, whereby we can fuse two interfaces $\text{CFT}_1|\text{CFT}_2$ and $\text{CFT}_2|\text{CFT}_3$ to get an interface $\text{CFT}_1|\text{CFT}_3$, or in terms of operators, $\mcO_{12}\circ\mcO_{23}=\mcO_{13}$.  If we fuse two interfaces which obey the consistency conditions, then the result will be a linear combination of consistent interfaces with nonnegative integer coefficients.  In the case\footnote{Sometimes conformal interfaces between copies of the same theory are called conformal defects.  In this paper we mostly use the term interface, but will occasionally use defect in this sense as well.} $\text{CFT}_1\sim\text{CFT}_2$, we always have the trivial interface, which acts as the identity in the fusion algebra.  This also allows us to make sense of the notion of whether a particular interface is invertible or not (most conformal interfaces are not invertible).

Part of the impetus to study conformal interfaces is that many physically interesting aspects of conformal field theory can be phrased in terms of interfaces and fusion.  For example, for any global symmetry present in the CFT, there will exist an invertible topological interface which implements that symmetry.  Fusion of these symmetry interfaces of course realize the symmetry group multiplication, and in general the conformal defects from a theory to itself form an algebraic structure known as the {\it{defect monoid}}~\cite{Bachas:2012bj}.  As another example, if two theories are connected by marginal deformation, then there exists an invertible interface (actually a family of them since we can combine it with a symmetry on either end of the deformation) which implements the adiabatic transformation of states and operators under the deformation.  If two theories are connected by a renormalization group flow, there will be an interface connecting the UV CFT with the IR CFT~\cite{brunner07a,gaiotto12,Becker:2017}, and by fusing such an interface with a boundary condition in the UV we can address the in general difficult problem of determining the fate of a given boundary condition under RG flow.  If $\text{CFT}_2$ is an orbifold of $\text{CFT}_1$, then there will be a conformal interface between them which implements the isomorphism between the untwisted sector of the orbifold and the invariant states of the parent.  Interfaces can also be used to encode T-duality and order-disorder dualities \cite{Frohlich09gb, Elitzur:2013ut}.

In this paper we construct a large class of Cardy-consistent boundary conditions in the $c=2$ product theories $(S^1)^2$, $S^1\times(S^1/\Z_2)$, and $(S^1/\Z_2)^2$, and then apply the unfolding procedure (the inverse of the folding trick) to obtain consistent conformal interfaces between pairs of $c=1$ theories which are either $S^1$ or $S^1/\Z_2$.  For the $(S^1)^2$ boundaries and $S^1|S^1$ interfaces, this is a review of the work of~\cite{bachas02,bachas07} (see also~\cite{fuchs07} for an alternative approach to constructing these interfaces).  For the orbifold theories these are new results, applying techniques that were developed by~\cite{affleck} to find conformal boundaries in the $S^1/\Z_2$ theory.  In particular, we find that in the $(S^1/\Z_2)^2$ case, for certain boundary state parameters we have to add particular combinations of twisted sector states in order to maintain Cardy consistency.  We then compute the full fusion algebra of these interfaces, and identify several particular interfaces of physical interest.  Our work aims to give a more complete picture of the spectrum of interfaces between the $c=1$ CFTs. Taken together, \cite{bachas07, fuchs07} and this note provide a more complete picture of the possible interfaces between theories in the space of $c=1$ CFTs~\cite{Ginsparg:1988ui}.

In section~\ref{sec:BCFT1}, we review the construction of consistent conformal boundary states in the $S^1$ and $S^1/\Z_2$ theories.  In section~\ref{sec:ProductBS}, we construct a large class of Cardy consistent boundary states in the $c=2$ product theories.  Most of the details of the annulus computations which are needed to establish Cardy consistency are relegated to appendix \ref{app:Annulus}.  Section \ref{sec:ConformalInterfaces} applies the unfolding trick to obtain the explicit conformal interface operators, and section \ref{sec:Fusion} computes the fusion algebra (with most of the details in appendix \ref{app:Fusion}) and identifies the interesting interfaces.  Finally, we discuss some possible future directions in section \ref{sec:Conclusions}.

\section{Boundary states in the circle and the orbifold}\label{review}
\label{sec:BCFT1}

In this section we review the boundary conformal field theory (BCFT) for the compact free boson at radius $R$ and its $\Z_2$ reflection orbifold, following~\cite{bachas07,affleck} using primarily the notation and conventions of~\cite{bachas07}.

\subsection{Circle theory}

We'll start with the compact free boson on a circle of radius $R$, which has mode expansion
\be
\label{eq:UntwistedModeExpansion}
X(\tau,\s)=\widehat{x}_0+\frac{\widehat{N}}{2R}\tau+\widehat{M}R\s+\sum_{n=1}^\infty\frac{i}{2\sqrt{n}}\ls a_ne^{-in(\tau+\s)}+\wta_ne^{-in(\tau-\s)}-a_n^\dagger e^{in(\tau+\s)}-\wta_n^\dagger e^{in(\tau-\s)}\rs,
\ee
where $a_n$ and $\wta_n$ are lowering operators and $a_n^\dagger=a_{-n}$, $\wta_n^\dagger=\wta_{-n}$ are the corresponding raising operators, $\widehat{N}$, $\widehat{M}$ are momentum and winding operators, and $\widehat{x}_0$ is the zero-mode operator, and we have commutation relations\footnote{Note that these differ from some standard conventions in string theory.  Also, we are essentially taking $\al'=1/2$.  The normalization for $\widehat{\widetilde{x}}_0$ is also convention; with our choice its eigenvalue has period $2\pi/R$.}
\be
\ls a_n,a_m^\dagger\rs=\ls\wta_n,\wta_n^\dagger\rs=\d_{n+m,0},\quad\ls\widehat{x}_0,\widehat{N}\rs=iR,\quad\ls\widehat{\widetilde{x}}_0,\widehat{M}\rs=-\frac{i}{R},
\ee
in which we have also introduced the operator $\widehat{\widetilde{x}}_0$ which is conjugate to winding number $\widehat{M}$ in the same way that $\widehat{x}_0$ is conjugate to momentum number $\widehat{N}$.  Generally, where confusion is possible, a hat an a symbol means that it's an operator; the same symbol without a hat is just a number, typically an eigenvalue of the corresponding operator.

The Hamiltonian of this theory is
\be
H=\frac{\widehat{N}^2}{4R^2}+\widehat{M}^2R^2+\sum_{n=1}^\infty n\lp a_n^\dagger a_n+\wta_n^\dagger\wta_n\rp-\frac{1}{12}.
\ee
The states in the Hilbert space are constructed by acting with raising operators on lowest weight states labeled by their $\widehat{N}$ and $\widehat{M}$ eigenvalues,
\be
|N,M\rangle=e^{i\frac{N}{R}\widehat{x}_0}e^{-iMR\widehat{\widetilde{x}}_0}|0\rangle.
\ee

We want to study the theory on the upper half-plane, with boundary conditions along the real axis.  In order for the conformal Ward identity to hold in the presence of the boundary condition~\cite{cardy}, the holomorphic and antiholomorphic stress tensors must agree at the boundary,
\be
T(z)=\widetilde{T}(\bz)|_{z=\bz}.
\ee
In the boundary state formalism, we formally construct a state (not in the Hilbert space of the bulk theory) $|B\rangle\rangle$ which satisfies
\be
\label{eq:BStateModes}
\lp L_n-\widetilde{L}_{-n}\rp|B\rangle\rangle=0,\qquad\forall n\in\Z.
\ee
There is a particularly easy way to achieve (\ref{eq:BStateModes}) in the circle theory with stress tensor $T(z)=-2:\p X(z)\p X(z):$, namely by demanding $\p X(z)=\pm\ov{\p}X(\bz)|_{z=\bz}$, corresponding to one of the following sets of conditions
\be
\label{eq:NDBCs}
\lp a_n+\wta_{-n}\rp|N\rangle\rangle=\widehat{N}|N\rangle\rangle=0,\qquad\lp a_n-\wta_{-n}\rp|D\rangle\rangle=\widehat{M}|D\rangle\rangle=0,
\ee
called Neumann or Dirichlet boundary conditions respectively.  A basis of formal states that satisfy these conditions is given by
\be
\label{eq:S1Ishibashi}
\lp\prod_{n=1}^\infty e^{-a_n^\dagger\wta_n^\dagger}\rp|0,M\rangle,\qquad\mathrm{or}\qquad\lp\prod_{n=1}e^{a_n^\dagger\wta_n^\dagger}\rp|N,0\rangle,
\ee
respectively, for any choice of $M$ or $N$.

It turns out that a better choice of basis is given by the Fourier transforms with particular normalization coefficients,
\bea
|N(\al)\rangle\rangle &=& \sqrt{R}\lp\prod_{n=1}^\infty e^{-a_n^\dagger\wta_n^\dagger}\rp\sum_{M\in\Z}e^{-iM\al}|0,M\rangle,\\
|D(\beta)\rangle\rangle &=& \frac{1}{\sqrt{2R}}\lp\prod_{n=1}e^{a_n^\dagger\wta_n^\dagger}\rp\sum_{N\in\Z}e^{-iN\beta}|N,0\rangle,
\eea
for Neumann and Dirichlet respectively.  Here $\al$ and $\beta$ are parameters with period $2\pi$, which we may take to be in the range $0\le\al,\beta<2\pi$.  They are related to eigenvalues of $\widehat{\wtx}_0$ or $\widehat{x}_0$ respectively, or more appropriately to the well-defined operators $\exp(-iR\widehat{\wtx}_0)$ or $\exp(i\widehat{x}_0/R)$,
\be
e^{-iR\widehat{\wtx}_0}|N(\al)\rangle\rangle=e^{i\al}|N(\al)\rangle\rangle,\qquad e^{\frac{i}{R}\widehat{x}_0}|D(\beta)\rangle\rangle=e^{i\beta}|D(\beta)\rangle\rangle.
\ee

The normalizations of these boundary states are chosen so that they satisfy the Cardy condition. If we compute the annulus diagram for any pair of allowed boundary conditions $B_i$ and $B_j$, 
\be
A_{ij}=\langle\langle B_i|q^H|B_j\rangle\rangle,
\ee
where the parameter $q=e^{-T}$, and if we perform a modular transformation to rewrite it in terms of $\wtq=e^{-\pi^2/T}$, then the result should have an expansion in terms of states with positive integer coefficients.  Moreover, if $i=j$, then the vacuum state (proportional to $\widetilde{q}^{-c/12}$) should appear with coefficient one, while for $i\ne j$ it should have coefficient zero.

For example, in the circle theory one can compute that
\bea
\label{eq:ANN}
A_{N(\al_1),N(\al_2)} &=& \lp R\sum_{M\in\Z}e^{-iM\lp\al_2-\al_1\rp}q^{M^2R^2}\rp\frac{1}{\eta(q^2)},\\
A_{N(\al_1),D(\beta_2)} &=& \sqrt{\frac{\eta(q^2)}{\vartheta_2(q^2)}},\\
\label{eq:ADD}
A_{D(\beta_1),D(\beta_2)} &=& \lp\frac{1}{2R}\sum_{N\in\Z}e^{-iN\lp\beta_2-\beta_1\rp}q^{\frac{N^2}{4R^2}}\rp\frac{1}{\eta(q^2)},
\eea
where our theta function conventions can be found in Appendix \ref{app:ThetaFunctions}, and the details of this computations (and others below) are relegated to Appendix \ref{app:Annulus}.  Performing the modular transformation, we find
\bea
\label{eq:ATNN}
A_{N(\al_1),N(\al_2)} &=& \frac{1}{\eta(\wtq^2)}\sum_{M\in\Z}\wtq^{\frac{1}{R^2}\lp M+\frac{\Delta\al}{2\pi}\rp^2},\\
\label{eq:ATND}
A_{N(\al_1),D(\beta_2)} &=& \sqrt{\frac{\eta(\wtq^2)}{\vartheta_4(\wtq^2)}},\\
\label{eq:ATDD}
A_{D(\beta_1),D(\beta_2)} &=& \frac{1}{\eta(\wtq^2)}\sum_{N\in\Z}\wtq^{4R^2\lp N+\frac{\Delta\beta}{2\pi}\rp^2},
\eea
with $\Delta\al=\al_2-\al_1$, $\Delta\beta=\beta_2-\beta_1$.  In this form it is easily verified that these boundary conditions are all mutually Cardy-consistent.  These are not the most general boundary states for the compact free boson theories, but they do provide a large class.  Specifically, these boundary states are the ones that preserve a $\U(1)$ symmetry (standard translation symmetry in the case of Neumann boundaries, or dual translations for Dirichlet boundaries), but there are other possibilities which do not preserve any continuous symmetry~\cite{Gaberdiel:2001zq,Janik:2001hb}.  Note that the boundary states (\ref{eq:S1Ishibashi}) (or any finite linear combinations thereof) are not even Cardy self-consistent.

\subsection{Orbifold theory}

This section follows Affleck and Oshikawa~\cite{affleck} closely.

Now we move on to the theory on $S^1/\Z_2$.  In this case the bulk theory has an untwisted sector, consisting of states of the circle theory which are invariant under the $\Z_2$ reflection, and a twisted sector, where fields are only periodic up to the action of the reflection and we again keep only invariant states.  In the untwisted sector we again use the mode expansion (\ref{eq:UntwistedModeExpansion}), and the $\Z_2$ flips the signs of all the operators appearing in the expansion (and also $\widehat{\widetilde{x}}_0$).  The lowest weight states transform as $|N,M\rangle\rr|-N,-M\rangle$, and we can build invariant untwisted states by taking sums of states from the parent theory, for example $(|N,M\rangle+|-N,-M\rangle)$, or $(a_{-1}|N,M\rangle-a_{-1}|-N,-M\rangle)$.

In the twisted sector, the mode expansion becomes
\be
X=\widehat{x}_0^t+\sum_{r\in\Z+\hlf,r>0}\frac{i}{2\sqrt{r}}\ls a_r^te^{-ir(\tau+\s)}+\wta_r^te^{-ir(\tau-\s)}-a_r^{t\,\dagger}e^{ir(\tau+\s)}-\wta_r^{t\,\dagger}e^{ir(\tau-\s)}\rs.
\ee
The superscripts $t$ are to remind us that these are operators acting only on the twisted sector of the Hilbert space (the operators without a superscript act only in the untwisted sector).  The eigenvalue zero mode operator $\widehat{x}_0^t$ must sit at one of the fixed points of the reflection, i.e.\ we must have either $x_0^t=0$ or $x_0^t=\pi R$, and the lowest weight states in the twisted sector are labeled by this fixed point, $|0\rangle^t$, $|\pi R\rangle^t$, and they are left invariant by the $\Z_2$, while the oscillators are flipped.  The Hamiltonian acting in this sector is
\be
H^t=\sum_{r\in\Z+\hlf,r>0}r\lp a_r^{t\,\dagger}a_r^t+\wta_r^{t\,\dagger}\wta_r^t\rp+\frac{1}{24}.
\ee

The simplest way to try and build boundary states of the orbifold theory is to take boundary states proportional to invariant combinations of boundary states in the parent theory.  It's not hard to verify that the $\Z_2$ action will flip the sign of the parameter in the boundary state, i.e.\ $|N(\al)\rangle\rangle\rr|N(-\al)\rangle\rangle$ or $|D(\beta)\rangle\rangle\rr|D(-\beta)\rangle\rangle$, so possible invariant combinations would be
\bea
|N_O(\al)\rangle\rangle &=& g_N(\al)\lp|N(\al)\rangle\rangle+|N(-\al)\rangle\rangle\rp,\\
|D_O(\beta)\rangle\rangle &=& g_D(\beta)\lp|D(\beta)\rangle\rangle+|D(-\beta)\rangle\rangle\rp,
\eea
where we have left some arbitrary normalization constants $g$, and without loss of generality we can assume $0\le\al,\beta\le\pi$.  To fix the normalizations, we must check the Cardy condition.  Computing the annulus diagrams and performing the modular transformations, we have
\bea
A_{N_O(\al_1),N_O(\al_2)} &=& \frac{2g_N(\al_1)g_N(\al_2)}{\eta(\wtq^2)}\lp\sum_{M\in\Z}\wtq^{\frac{1}{R^2}\lp M+\frac{\al_2-\al_1}{2\pi}\rp^2}+\sum_{M\in\Z}\wtq^{\frac{1}{R^2}\lp M+\frac{\al_1+\al_2}{2\pi}\rp^2}\rp,\\
A_{N_O(\al_1),D_O(\beta_2)} &=& 4g_N(\al_1)g_D(\beta_2)\sqrt\frac{\eta(\wtq^2)}{\vartheta_4(\wtq^2)},\\
A_{D_O(\beta_1),D_O(\beta_2)} &=& \frac{2g_D(\beta_1)g_D(\beta_2)}{\eta(\wtq^2)}\lp\sum_{N\in\Z}\wtq^{4R^2\lp N+\frac{\beta_2-\beta_1}{2\pi}\rp^2}+\sum_{N\in\Z}\wtq^{4R^2\lp N+\frac{\beta_1+\beta_2}{2\pi}\rp^2}\rp.
\eea
Considering the diagonal results, we learn that Cardy self-consistency divides these states into two classes.  For the Neumann states, we get the requirement that
\be
g_N(\al)^2=\left\{\begin{matrix}\hlf, & 0<\al<\pi,\\ \frac{1}{4}, & \al=0\ \mathrm{or}\ \pi,\end{matrix}\right.
\ee
while for Dirichlet states we find
\be
g_D(\beta)^2=\left\{\begin{matrix}\hlf, & 0<\beta<\pi,\\ \frac{1}{4}, & \beta=0\ \mathrm{or}\ \pi.\end{matrix}\right.
\ee
For the generic parameter choices, $0<\al,\beta<\pi$, then we take $g_N(\al)$ or $g_D(\beta)$ equal to $1/\sqrt{2}$, and those states will all be mutually consistent.  We will call these the generic states.  However when the parameters take values at the end points, it is not possible to satisfy both mutual consistency and consistency with generic states.

There are other possibilities to construct boundary states in the orbifold theory however, by building them on top of twisted sector ground states.  There are two possibilities to satisfy Dirichlet boundary conditions on such a state (we will put a subscript $0$ on a parameter like $\al$ or $\beta$ when we want to emphasize that it must take on a fixed point value of either $0$ or $\pi$),
\be
|D_O(\beta_0)\rangle\rangle^t=\lp\prod_{r\in\Z+\hlf,r>0}e^{a_r^{t\,\dagger}\wta_r^{t\,\dagger}}\rp|\beta_0 R\rangle^t.
\ee
where $\beta_0=0$ or $\pi$, and similarly two possibilities for Neumann boundary conditions,
\be
\lp\prod_{r\in\Z+\hlf,r>0}e^{-a_r^{t\,\dagger}\wta_r^{t\,\dagger}}\rp|\beta_0 R\rangle^t.
\ee
For these last two, it is more useful to take combinations $|N_O(\al_0)\rangle\rangle^t$, $\al_0=0,\pi$, defined as
\bea
|N_O(0)\rangle\rangle^t &=& \frac{1}{\sqrt{2}}\lp\prod_{r\in\Z+\hlf,r>0}e^{-a_r^{t\,\dagger}\wta_r^{t\,\dagger}}\rp\lp|0\rangle^t+|\pi R\rangle^t\rp,\\
|N_O(\pi)\rangle\rangle^t &=& \frac{1}{\sqrt{2}}\lp\prod_{r\in\Z+\hlf,r>0}e^{-a_r^{t\,\dagger}\wta_r^{t\,\dagger}}\rp\lp|0\rangle^t-|\pi R\rangle^t\rp.
\eea
These choices are made so that the states are eigenstates of the twisted sector analogs of the operators $\widehat{x}_0$ and $\widehat{\wtx}_0$,
\be
\label{eq:ZMEigenvaluesTS}
\widehat{x}_0^t|D_O(\beta_0)\rangle\rangle^t=R\beta_0|D_O(\beta_0)\rangle\rangle^t,\quad\widehat{\wtx}_0^t|N_O(\al_0)\rangle\rangle^t=\frac{\al_0}{R}|N_O(\al_0)\rangle\rangle^t.
\ee
We haven't yet picked the correct linear combinations for these states.  To fix these we must solve the Cardy consistency conditions.

By computing annulus diagrams (see Appendix \ref{app:OrbifoldAnnulus}), it can be shown that we can get mutually consistent (and also consistent with the generic parameter untwisted boundary states) states by taking certain combinations with the problematic endpoint parameter states from above.  Summarizing those results, we find a large set of mutually consistent boundary states given by
\bea
\label{eq:OrbB1}
|N_O(\al)\rangle\rangle &=& \frac{1}{\sqrt{2}}|N(\al)\rangle\rangle+\frac{1}{\sqrt{2}}|N(-\al)\rangle\rangle,\qquad 0<\al<\pi,\\
|N_O(\al_0,\pm)\rangle\rangle &=& \frac{1}{\sqrt{2}}|N(\al_0)\rangle\rangle\pm 2^{-\frac{1}{4}}
|N_O(\al_0)\rangle\rangle^t,\qquad\al_0\in\{0,\pi\},\\
|D_O(\beta)\rangle\rangle &=& \frac{1}{\sqrt{2}}|D(\beta)\rangle\rangle+\frac{1}{\sqrt{2}}|D(-\beta)\rangle\rangle,\qquad 0<\beta<\pi,\\
\label{eq:OrbB4}
|D_O(\beta_0,\pm)\rangle\rangle &=& \frac{1}{\sqrt{2}}|D(\beta_0)\rangle\rangle\pm 2^{-\frac{1}{4}}|D_O(\beta_0)\rangle\rangle^t,\qquad \beta_0\in\{0,\pi\}.
\eea
Each of the Neumann and Dirichlet states satisfy (\ref{eq:NDBCs}), where now either $n\in\Z$ for untwisted sector oscillators, or $n\in\Z+\hlf$ for twisted sector oscillators (and of course there are no $\widehat{M}$ or $\widehat{N}$ operators in the twisted sector).  We have also imposed one extra condition, which is that the boundary states are eigenstates of the operator (understood now to be acting in both untwisted and twisted sectors, where we would add a superscript $t$) $\exp(i\widehat{x}_0/R)$ (for Dirichlet) or $\exp(iR\widehat{\wtx}_0)$ (for Neumann) with eigenvalues $\pm 1$.  The Cardy conditions alone would not prevent one from taking combinations like
\be
\frac{1}{\sqrt{2}}|D(0)\rangle\rangle\pm 2^{-\frac{1}{4}}|D_O(\pi)\rangle\rangle^t.
\ee
These states would presumably violate other consistency conditions (sewing conditions), but without doing a full analysis of those broader constraints, a good proxy seems to be the natural idea that we should take eigenstates of the appropriate zero-mode operators.

Finally, let's note that these boundary states satisfy various threshold conditions as the generic states approach endpoint values, i.e.\
\be
\label{eq:O1Threshold}
\lim_{\al\rr\al_0}|N_O(\al)\rangle\rangle=|N_O(\al_0,+)\rangle\rangle+|N_O(\al_0,-)\rangle\rangle,\quad\lim_{\beta\rr\beta_0}|D_O(\beta)\rangle\rangle=|D_O(\beta_0,+)\rangle\rangle+|D_O(\beta_0,-)\rangle\rangle.
\ee

\section{Boundary states in the product theories}
\label{sec:ProductBS}

In this section we would like to construct as many boundary states as possible for bulk CFTs of the form CFT$_1\otimes$CFT$_2$, wher each of CFT$_1$ and CFT$_2$ is either the $S^1$ or the $S^1/\Z_2$ theory.  There are certain obvious boundary states obtained by simply taking tensor products of the boundary states found in section \ref{sec:BCFT1}.  Following~\cite{bachas07}, we will also look for more general rotated branes in the product theories.


\subsection{Boundary states in the $(S^1)^2$ theory}

First we consider the product of two circle theories, with radii $R_1$ and $R_2$ respectively, and construct a large set of consistent conformal boundary conditions following~\cite{bachas07}.

The most obvious examples we can construct are factorized boundary states,
\be
|N(\al_1)\rangle\rangle_1\otimes|N(\al_2)\rangle\rangle_2,\quad|N(\al_1)\rangle\rangle_1\otimes|D(\beta_2)\rangle\rangle_2,\quad|D(\beta_1)\rangle\rangle_1\otimes|N(\al_2)\rangle\rangle_2,\quad|D(\beta_1)\rangle\rangle_1\otimes|D(\beta_2)\rangle\rangle_2.
\ee
The fact that these are all mutually Cardy consistent follows from the consistency in the $S^1$ theories.  It turns out, however, that there are many more possibilities we can construct.  The key observation is that the stress tensor in the product theory
\be
T(z)=-2\lp:\p X^1(z)\p X^1(z):+:\p X^2(z)\p X^2(z):\rp,
\ee
is invariant under rotations of $X^1$ and $X^2$.  This means that instead of solving (\ref{eq:BStateModes}) by separately imposing either Neumann or Dirichlet conditions on $X^1$ and $X^2$, we can instead build rotated combinations $Y^1=\cos\vartheta X^1+\sin\vartheta X^2$, $Y^2=-\sin\vartheta X^1+\cos\vartheta X^2$, and impose conditions on these, say Neumann on $Y^1$ and Dirichlet on $Y^2$.

For the oscillators, it is easy to construct a boundary state with these properties.  Indeed, if we let $b_n^1$, $\widetilde{b}_n^1$, $b_n^2$, and $\widetilde{b}_n^2$ be the oscillators for $Y^1$ and $Y^2$, i.e.\
\be
\lp\begin{matrix}b_n^1 \\ b_n^2\end{matrix}\rp=\mathcal{R}(\vartheta)\lp\begin{matrix}a_n^1 \\ a_n^2\end{matrix}\rp=\lp\begin{matrix}\cos\vartheta a_n^1+\sin\vartheta a_n^2 \\ -\sin\vartheta a_n^1+\cos\vartheta a_n^2\end{matrix}\rp,
\ee
where
\be
\mathcal{R}(\vartheta)=\lp\begin{matrix}\cos\vartheta & \sin\vartheta \\ -\sin\vartheta & \cos\vartheta\end{matrix}\rp,
\ee
is the rotation matrix, then we simply need to act on a lowest weight state with an operator
\be
\prod_{n=1}^\infty e^{-b_n^{1\,\dagger}\widetilde{b}_n^{1\,\dagger}+b_n^{2\,\dagger}\widetilde{b}_n^{2\,\dagger}}=\prod_{n=1}^\infty e^{a_n^{\dagger\,T}\cdot S^{(+)}\cdot\wta_n^\dagger},
\ee
where we use the matrix
\be
S^{(+)}(\vartheta)=\mathcal{R}(\vartheta)^T\lp\begin{matrix} -1 & 0 \\ 0 & 1\end{matrix}\rp\mathcal{R}(\vartheta)=\lp\begin{matrix} -\cos 2\vartheta & -\sin 2\vartheta \\ -\sin 2\vartheta & \cos 2\vartheta\end{matrix}\rp.
\ee
We will often omit the $\vartheta$ argument of $S^{(+)}$, leaving it implicit.

Turning now to the zero mode pieces of $Y^1$ and $Y^2$, we find the conditions
\be
\lp\cos\vartheta\frac{\widehat{N}^1}{R_1}+\sin\vartheta\frac{\widehat{N}^2}{R_2}\rp|B\rangle\rangle=\lp -R_1\sin\vartheta\widehat{M}^1+R_2\cos\vartheta\widehat{M}^2\rp|B\rangle\rangle=0.
\ee
Since $\widehat{N}^1$, $\widehat{N}^2$, $\widehat{M}^1$, and $\widehat{M}^2$ have only integer eigenvalues, we find the restriction that there must be relatively prime integers $k_1$ and $k_2$ such that
\be
\frac{R_1}{R_2}\tan\vartheta=\frac{k_2}{k_1},
\ee
in which case we set $N^1/N^2=-k_2/k_1$ and $M^1/M^2=k_1/k_2$.  Then a set of states obeying our rotated ND boundary conditions are thus
\be
|B^{(+)}_{k_1,k_2}(\al;\beta)\rangle\rangle=g_{k_1,k_2}(\al,\beta)\lp\prod_{n=1}^\infty e^{a_n^{\dagger\,T}\cdot S^{(+)}\cdot\wta_n^\dagger}\rp\sum_{N,M\in\Z}e^{-iM\al-iN\beta}|-k_2N,k_1M\rangle_1\otimes|k_1N,k_2M\rangle_2,
\ee
where as before we have $0\le\al,\beta<2\pi$.  The superscript $(+)$ indicates that we use the matrix $S^{(+)}$ in the exponential, and serves to distinguish from another class of boundary states that we will introduce below.  Note that there is a small redundancy in these states,
\be
\label{eq:Redundancy}
|B^{(+)}_{k_1,k_2}(\al;\beta)\rangle\rangle=|B^{(+)}_{-k_1,-k_2}(-\al;-\beta)\rangle\rangle.
\ee
Note also that we can easily express the entries of the matrix $S^{(+)}$ directly in terms of $k_1$ and $k_2$ by using
\be
\vartheta=\tan^{-1}\lp\frac{k_2R_2}{k_1R_1}\rp,\qquad\cos 2\vartheta=\frac{k_1^2R_1^2-k_2^2R_2^2}{k_1^2R_1^2+k_2^2R_2^2},\qquad\sin 2\vartheta=\frac{2k_1k_2R_1R_2}{k_1^2R_1^2+k_2^2R_2^2}.
\ee
Finally, observe that our original factorized states are special cases,
\be
|N(\al_1)\rangle\rangle_1\otimes|D(\beta_2)\rangle\rangle_2=|B^{(+)}_{1,0}(\al;\beta)\rangle\rangle,\qquad|D(\beta_1)\rangle\rangle_1\otimes|N(\al_2)\rangle\rangle_2=|B^{(+)}_{0,1}(\al;-\beta)\rangle\rangle.
\ee
However, since for many later purposes the factorized cases behave qualitatively different from the non-factorized cases, we will assume, unless stated otherwise, that $k_1$ and $k_2$ are both non-zero, and we will write the factorized boundary states as explicit products.

To check that these are mutually consistent boundary states, and to fix the normalization constants, we compute the relevant annulus diagrams.  The details are relegated to Appendix \ref{app:T2Ann}.  If $(k_1',k_2')=(k_1,k_2)$ (or something similar if $(k_1',k_2')=(-k_1,-k_2)$), we have
\be
\label{eq:ANDND1}
A_{B^{(+)}_{k_1,k_2}(\al;\beta),B^{(+)}_{k_1',k_2'}(\al';\beta')}=\frac{2R_1R_2gg'}{k_1^2R_1^2+k_2^2R_2^2}\frac{1}{\eta(\widetilde{q}^2)^2}\sum_{M,N\in\Z}\widetilde{q}^{\frac{4R_1^2R_2^2\lp N-\frac{\Delta\beta}{2\pi}\rp^2+\lp M-\frac{\Delta\al}{2\pi}\rp^2}{k_1^2R_1^2+k_2^2R_2^2}},
\ee
and otherwise we have
\be
\label{eq:ANDND2}
A_{B^{(+)}_{k_1,k_2}(\al;\beta),B^{(+)}_{k_1',k_2'}(\al';\beta')}=\frac{2R_1R_2gg'\lp k_1k_2'-k_1'k_2\rp}{\sqrt{k_1^2R_1^2+k_2^2R_2^2}\sqrt{k_1^{\prime\,2}R_1^2+k_2^{\prime\,2}R_2^2}}\frac{i\eta(\widetilde{q}^2)\widetilde{q}^{\,-\lp\frac{\Delta\vartheta}{\pi}\rp^2}}{\vartheta_1(\widetilde{q}^2,i\frac{\Delta\vartheta}{T})}.
\ee
These expressions actually are valid even in the factorized case if we allow some $k_a$ or $k_a'$ to vanish.  From these expressions we can check that these states are all mutually Cardy consistent provided we choose normalizations
\be
g_{k_1,k_2}(\al,\beta)=\sqrt{\frac{k_1^2R_1^2+k_2^2R_2^2}{2R_1R_2}}.
\ee

Another set of boundary states can be obtained from these by applying T-duality.  On a single compact boson, T-duality acts as
\be
R\rightarrow\frac{1}{2R},\quad\wta_n\rightarrow -\wta_n,\quad\wta_n^\dagger\rightarrow -\wta_n^\dagger,\quad\mathrm{and}\quad\left|N,M\right\rangle\rightarrow\left|M,N\right\rangle.
\ee
Applying this map to $X^1$ only in our states $|B^{(+)}_{k_1,k_2}(\al;\beta)\rangle\rangle$, we get new states~\cite{bachas07}
\begin{multline}
|B^{(-)}_{k_1,k_2}(\al;\beta)\rangle\rangle\\
=\sqrt{\frac{k_1^2+4k_2^2R_1^2R_2^2}{4R_1R_2}}\lp\prod_{n=1}^\infty e^{a^{\dagger\,T}\cdot S^{(-)}\cdot\wta^\dagger}\rp\sum_{N,M\in\Z}e^{-iM\al-iN\beta}|k_1M,-k_2N\rangle_1\otimes|k_1N,k_2M\rangle_2,
\end{multline}
where we have defined
\be
S^{(-)}=S^{(+)}\lp\begin{matrix}-1 & 0 \\ 0 & 1\end{matrix}\rp=\lp\begin{matrix}\cos 2\theta & -\sin 2\theta \\ \sin 2\theta & \cos 2\theta\end{matrix}\rp,
\ee
and
\be
\theta=\tan^{-1}\lp\frac{2k_2R_1R_2}{k_1}\rp.
\ee

These states include our original factorized NN and DD states,
\be
|N(\al_1)\rangle\rangle_1\otimes|N(\al_2)\rangle\rangle_2=|B^{(-)}_{0,1}(\al_2;-\al_1))\rangle\rangle,\qquad|D(\beta_1)\rangle\rangle_1\otimes|D(\beta_2)\rangle\rangle_2=|B^{(-)}_{1,0}(\beta_1;\beta_2)\rangle\rangle.
\ee
The fact that all these states are consistent with each other follows from T-duality.  To check the consistency with the rotated ND states, we compute one more annulus diagram,
\be
\label{eq:ANDDD}
A_{B^{(+)}_{k_1,k_2}(\al;\beta),B^{(-)}_{k_1',k_2'}(\al';\beta')}=\frac{1}{\sqrt{\eta(\widetilde{q}^2)\vartheta_4(\widetilde{q}^2)}}\sum_{L\in\Z}\widetilde{q}^{\frac{4R_1^2R_2^2\lp L+\frac{1}{2\pi}\lp k_2'\al-k_1'\beta-k_2\al'+k_1\beta'\rp\rp^2}{\lp k_1^2R_1^2+k_2^2R_2^2\rp\lp k_1^{\prime\,2}+4R_1^2R_2^2k_2^{\prime\,2}\rp}}.
\ee
This shows that all of these states are mutually Cardy consistent.

\subsection{Orbifold theories}
\label{subsec:OrbProdBS}

We would now like to find a set of consistent boundary conditions for $S^1\times (S^1/\Z_2)$ and for $(S^1/\Z_2)^2$ theories.  We will proceed in the same way that we did for the $S^1/\Z_2$ case, by first trying boundary states that are simply invariant combinations of $(S^1)^2$ boundary states, and then augmenting as necessary by twisted sector states.

We'll start with $S^1\times (S^1/\Z_2)$, a circle theory of radius $R_1$ and an orbifold of a circle with radius $R_2$.  In the parent theory we would have the ND and DD states found in the previous subsection.  The reflection in theory 2 acts on these by
\be
|B^{(\pm)}_{k_1,k_2}(\al;\beta)\rangle\rangle\rr|B^{(\pm)}_{k_1,-k_2}(\al;-\beta)\rangle\rangle.
\ee
Thus, we should consider the following invariant states
\be
|B^{(\pm)\,SO}_{k_1,k_2}(\al;\beta)\rangle\rangle=g^{(\pm)\,SO}_{k_1,k_2}(\al;\beta)\lp|B^{(\pm)}_{k_1,k_2}(\al;\beta)\rangle\rangle+|B^{(\pm)}_{k_1,-k_2}(\al;-\beta)\rangle\rangle\rp,
\ee
where $k_1$ and $k_2$ are relatively prime integers, $0\le\al,\beta<2\pi$, and we have two equivalences,
\be
|B^{(\pm)\,SO}_{k_1,k_2}(\al;\beta)\rangle\rangle\sim|B^{(\pm)\,SO}_{-k_1,-k_2}(-\al;-\beta)\rangle\rangle\sim|B^{(\pm)\,SO}_{k_1,-k_2}(\al;-\beta)\rangle\rangle.
\ee
We can use these equivalences to ensure that the $k_1$ and $k_2$ used to label our state are both non-negative.  The $SO$ superscripts indicate that theory 1 is the $S^1$ theory, while theory 2 is the orbifold $S^1/\Z_2$.  If we omit this superscript, then we mean the $S^1\times S^1$ boundary state.

Checking self-consistency, we learn that we would need to take $g^{(\pm)\,SO}_{k_1,k_2}(\al,\beta)=1/\sqrt{2}$ except when the two terms are equal by the equivalence (\ref{eq:Redundancy}).  This can only happen in the factorized case, when either $k_1=0$ and $\al\in\{0,\pi\}$, or $k_2=0$ and $\beta=\{0,\pi\}$.
In these cases we can easily repair the conflict between self-consistency and mutual consistency
by taking factorized states that include the Affleck-Oshikawa twisted sector states as a factor.  This leaves us with (in addition to the factorized states which are products of $S^1$ boundary states and generic parameter $S^1/\Z_2$ boundary states)
\be
|N(\al)\rangle\rangle_1\otimes|N_O(\beta_0,\e)\rangle\rangle_2,\quad|N(\al)\rangle\rangle_1\otimes|D_O(\beta_0,\e)\rangle\rangle_2,
\ee
and so on.  All of these states can then be verified to be consistent.

Next we look at the $(S^1/\Z_2)^2$ theories.  Proceeding in exactly the same fashion as above, we find states for generic parameters,
\begin{multline}
|B^{(\pm)\,OO}_{k_1,k_2}(\al;\beta)\rangle\rangle=\hlf\lp|B^{(\pm)}_{k_1,k_2}(\al;\beta)\rangle\rangle+|B^{(\pm)}_{k_1,-k_2}(-\al;\beta)\rangle\rangle\right.\\
\left. +|B^{(\pm)}_{k_1,-k_2}(\al;-\beta)\rangle\rangle+|B^{(\pm)}_{k_1,k_2}(-\al;-\beta)\rangle\rangle\rp.
\end{multline}
with equivalences
\be
|B^{(\pm)\,OO}_{k_1,k_2}(\al;\beta)\rangle\rangle\sim|B^{(\pm)\,OO}_{-k_1,-k_2}(-\al;-\beta)\rangle\rangle\sim|B^{(\pm)\,OO}_{k_1,-k_2}(\al;-\beta)\rangle\rangle\sim|B^{(\pm)\,OO}_{k_1,k_2}(-\al;-\beta)\rangle\rangle.
\ee
These are self-consistent states unless there is an equivalence between terms in the linear combination, which occurs in three situations,
\begin{itemize}
\item $k_2=0$, at least one of $\al\in\{0,\pi\}$ or $\beta\in\{0,\pi\}$,
\item $k_1=0$, at least one of $\al\in\{0,\pi\}$ or $\beta\in\{0,\pi\}$,
\item both $k_1$ and $k_2$ nonzero (by equivalence we can take them both positive), and both $\al\in\{0,\pi\}$ and $\beta\in\{0,\pi\}$.
\end{itemize}

For the first two cases, we can repair the consistency by simply taking factorized states, such as
\be
|N_O(\al)\rangle\rangle_1\otimes|D_O(\pi,\e')\rangle\rangle_2,\qquad\mathrm{or}\qquad|N_O(0,\e)\rangle\rangle_1\otimes|N_O(\pi,\e')\rangle\rangle_2,
\ee
etc, where $\e$ and $\e'$ are $\pm$.  We can easily verify that these are Cardy consistent with the generic states.

For the last case, with both $k_1$ and $k_2$ nonzero, we can also repair consistency by taking particular combinations of twisted-twisted states $|\g_0\rangle_1^t\otimes|\d_0\rangle_2^t$.  Which combinations we take are not entirely determined by Cardy consistency, so we need another criterion.  Guided by the $S^1/\Z_2$ case, we will demand that the boundary states be eigenstates of some zero-mode operators.  But which operators?  We must examine the symmetrized states and check whether they are eigenstates of any zero-mode operators.  This can be determined by simply rotating the answer for the factorized ND states.  We find that $|B^{(+)}_{k_1,k_2}(\al;\beta)\rangle\rangle$ is an eigenstate of the operators
\be
\lp\cos\vartheta\widehat{\wtx}_0^1+\sin\vartheta\widehat{\wtx}_0^2\rp\qquad\mathrm{and}\qquad\lp -\sin\vartheta\widehat{x}_0^1+\cos\vartheta\widehat{x}_0^2\rp,
\ee
with eigenvalues
\be
-\frac{\al}{\sqrt{k_1^2R_1^2+k_2^2R_2^2}}\qquad\mathrm{and}\qquad\frac{R_1R_2\beta}{\sqrt{k_1^2R_1^2+k_2^2R_2^2}},
\ee
respectively.  Or, rephrasing more usefully,
\be
e^{-ik_1R_1\widehat{\wtx}_0^1-ik_2R_2\widehat{\wtx}_0^2}|B^{(+)}_{k_1,\pm k_2}(\al;\beta)\rangle\rangle=e^{i\al}|B^{(+)}_{k_1,\pm k_2}(\al;\beta)\rangle\rangle,
\ee
\be
e^{-ik_2\frac{\widehat{x}_0^1}{R_1}+ik_1\frac{\widehat{x}_0^2}{R_2}}|B^{(+)}_{k_1,\pm k_2}(\al;\beta)\rangle\rangle=e^{i\beta}|B^{(+)}_{k_1,\pm k_2}(\al;\beta)\rangle\rangle.
\ee
Using the results (\ref{eq:ZMEigenvaluesTS}), we can determine how these operators (or rather their twisted sector analogs) act on the twisted-twisted states.  Taking an ordered basis
\be
\label{eq:TTBasis}
\left\{v_{00}=|0\rangle_1^t\otimes|0\rangle_2^t,\ v_{01}=|0\rangle_1^t\otimes|\pi R_2\rangle_2^t,\ v_{10}=|\pi R_1\rangle_1^t\otimes|0\rangle_2^t,\ v_{11}=|\pi R_1\rangle_1^t\otimes|\pi R_2\rangle_2^t\right\},
\ee
we can work out the action of these operators by exponentiating the action of the zero-mode operators,
\be
e^{-ik_1R_1\widehat{\wtx}_0^1-ik_2R_2\widehat{\wtx}_0^2}=\lp\begin{smallmatrix}\Pi_1\Pi_2 & \Pi_1(1-\Pi_2) & (1-\Pi_1)\Pi_2 & (1-\Pi_1)(1-\Pi_2) \\ \Pi_1(1-\Pi_2) & \Pi_1\Pi_2 & (1-\Pi_1)(1-\Pi_2) & (1-\Pi_1)\Pi_2 \\ (1-\Pi_1)\Pi_2 & (1-\Pi_1)(1-\Pi_2) & \Pi_1\Pi_2 & \Pi_1(1-\Pi_2) \\ (1-\Pi_1)(1-\Pi_2) & (1-\Pi_1)\Pi_2 & \Pi_1(1-\Pi_2) & \Pi_1\Pi_2\end{smallmatrix}\rp,
\ee
\be
e^{-ik_2\frac{\widehat{x}_0^1}{R_1}+ik_1\frac{\widehat{x}_0^2}{R_2}}=\lp\begin{smallmatrix}1 & 0 & 0 & 0 \\ 0 & (-1)^{k_1} & 0 & 0 \\ 0 & 0 & (-1)^{k_2} & 0 \\ 0 & 0 & 0 & (-1)^{k_1+k_2}\end{smallmatrix}\rp,
\ee
where we have defined
\be
\Pi_1=\frac{1+(-1)^{k_1}}{2},\qquad\Pi_2=\frac{1+(-1)^{k_2}}{2}.
\ee
The system of eigenvalues and eigenvectors thus depends on whether each of $k_1$ and $k_2$ is even or odd.  For each choice of $k_1$ and $k_2$, and each choice of $\al$ and $\beta$, we simply need to find the twisted sector state that has the same eigenvalues under the operators above.  The unit eigenvectors for these operators are tabulated in Table \ref{tab:vecs}.

Proceeding similarly for the $|B^{(-)}_{k_1,\pm k_2}\rangle\rangle$ states, we find that they are eigenstates of the operators
\be
e^{ik_1\frac{\widehat{x}_0^1}{R_1}-ik_2R_2\widehat{\wtx}_0^2}=\lp\begin{smallmatrix}\Pi_2 & 1-\Pi_2 & 0 & 0 \\ 1-\Pi_2 & \Pi_2 & 0 & 0 \\ 0 & 0 & (-1)^{k_1}\Pi_2 & (-1)^{k_1}(1-\Pi_2) \\ 0 & 0 & (-1)^{k_1}(1-\Pi_2) & (-1)^{k_1}\Pi_2\end{smallmatrix}\rp,
\ee
and
\be
e^{ik_2R_1\widehat{\wtx}_0^1+ik_1\frac{\widehat{x}_0^2}{R_2}}=\lp\begin{smallmatrix}\Pi_2 & 0 & 1-\Pi_2 & 0 \\ 0 & (-1)^{k_1}\Pi_2 & 0 & (-1)^{k_1}(1-\Pi_2) \\ 1-\Pi_2 & 0 & \Pi_2 & 0 \\ 0 & (-1)^{k_1}(1-\Pi_2) & 0 & (-1)^{k_1}\Pi_2\end{smallmatrix}\rp,
\ee
with eigenvalues $e^{i\al}$ and $e^{i\beta}$ respectively, where the matrix form above represents how they act on our basis (\ref{eq:TTBasis}) of twisted-twisted states.  We can again compute the unit eigenvectors of these matrices, and they are also listed in Table \ref{tab:vecs}.

\begin{table}
\centering
\begin{tabular}{|c|c|c|c|}
\hline
& $k_1$ and $k_2$ odd & $k_1$ odd, $k_2$ even & $k_1$ even, $k_2$ odd \\ \hline \hline
$|v^{(+)}_{k_1,k_2}(0;0)\rangle^t$ & $\tfrac{1}{\sqrt{2}}(v_{00}+v_{11})$ & $\tfrac{1}{\sqrt{2}}(v_{00}+v_{10})$ & $\tfrac{1}{\sqrt{2}}(v_{00}+v_{01})$ \\ \hline
$|v^{(+)}_{k_1,k_2}(0;\pi)\rangle^t$ & $\tfrac{1}{\sqrt{2}}(v_{01}+v_{10})$ & $\tfrac{1}{\sqrt{2}}(v_{01}+v_{11})$ & $\tfrac{1}{\sqrt{2}}(v_{10}+v_{11})$ \\ \hline
$|v^{(+)}_{k_1,k_2}(\pi;0)\rangle^t$ & $\tfrac{1}{\sqrt{2}}(v_{00}-v_{11})$ & $\tfrac{1}{\sqrt{2}}(v_{00}-v_{10})$ & $\tfrac{1}{\sqrt{2}}(v_{00}-v_{01})$ \\ \hline
$|v^{(+)}_{k_1,k_2}(\pi;\pi)\rangle^t$ & $\tfrac{1}{\sqrt{2}}(v_{01}-v_{10})$ & $\tfrac{1}{\sqrt{2}}(v_{01}-v_{11})$ & $\tfrac{1}{\sqrt{2}}(v_{10}-v_{11})$ \\ \hline \hline
$|v^{(-)}_{k_1,k_2}(0;0)\rangle^t$ & $\tfrac{1}{2}(v_{00}+v_{01}+v_{10}-v_{11})$ & $v_{00}$ & $\tfrac{1}{2}(v_{00}+v_{01}+v_{10}+v_{11})$ \\ \hline
$|v^{(-)}_{k_1,k_2}(0;\pi)\rangle^t$ & $\tfrac{1}{2}(v_{00}+v_{01}-v_{10}+v_{11})$ & $v_{01}$ & $\tfrac{1}{2}(v_{00}+v_{01}-v_{10}-v_{11})$ \\ \hline
$|v^{(-)}_{k_1,k_2}(\pi;0)\rangle^t$ & $\tfrac{1}{2}(v_{00}-v_{01}+v_{10}+v_{11})$ & $v_{10}$ & $\tfrac{1}{2}(v_{00}-v_{01}+v_{10}-v_{11})$ \\ \hline
$|v^{(-)}_{k_1,k_2}(\pi;\pi)\rangle^t$ & $\tfrac{1}{2}(v_{00}-v_{01}-v_{10}-v_{11})$ & $v_{11}$ & $\tfrac{1}{2}(v_{00}-v_{01}-v_{10}+v_{11})$ \\ \hline
\end{tabular}
\caption{The appropriate unit eigenvectors $|v^{(\eta)}_{k_1,k_2}(\al_0;\beta_0)\rangle^t$ for given $\eta$, $k_1$, $k_2$, $\al_0$, and $\beta_0$.}
\label{tab:vecs}
\end{table}

Let's define
\be
\Om^{(\pm)}=\hlf\lp\prod_{r\in\Z+\hlf,r>0}e^{a_r^{t\,\dagger\,T}\cdot S^{(\pm)}(\theta)\cdot\wta_r^{t\,\dagger}}\rp+\hlf\lp\prod_{r\in\Z+\hlf,r>0}e^{a_r^{t\,\dagger\,T}\cdot S^{(\pm)}(-\theta)\cdot\wta_r^{t\,\dagger}}\rp.
\ee
This provides the appropriate invariant combination of oscillator modes to act on the twisted sector zero-mode states to implement the desired boundary conditions.

With this notation, the non-generic non-factorized $(S^1/\Z^2)^2$ boundary states are all given by taking the invariant untwisted sector piece and adding to it the appropriate twisted sector eigenvector, with normalization fixed by the Cardy condition,
\be
|B^{(\pm)\,OO}_{k_1,k_2}(\al_0;\beta_0;\e)\rangle\rangle=\hlf\lp|B^{(\pm)}_{k_1,k_2}(\al_0;\beta_0)\rangle\rangle+|B^{(\pm)}_{k_1,-k_2}(\al_0;\beta_0)\rangle\rangle\rp+\sqrt{2}\e\Om^{(\pm)}|v^{(\pm)}(\al_0;\beta_0)\rangle^t,
\ee
where $|v^{(\pm)}(\al_0;\beta_0)\rangle^t$ is determined from Table \ref{tab:vecs}.
The full set of states can be checked to be all mutually Cardy consistent.

We also have threshold decompositions as generic states approach one of these special states, i.e.\
\be
\lim_{\al\rr\al_0,\beta\rr\beta_0}|B^{(\pm)\,OO}_{k_1,k_2}(\al;\beta)\rangle\rangle=|B^{(\pm)\,OO}_{k_1,k_2}(\al_0;\beta_0;+)\rangle\rangle+|B^{(\pm)\,OO}_{k_1,k_2}(\al_0;\beta_0;-)\rangle\rangle.
\ee

\section{Conformal interfaces}
\label{sec:ConformalInterfaces}

In this section we use the unfolding trick to~\cite{affleck,bachas02} convert our boundary states from section \ref{sec:ProductBS} into conformal interfaces between different combinations of $S^1$ and $S^1/\Z_2$ theories.

Given two two-dimensional CFTs separated by a one-dimensional conformal interface (which we denote by $\text{CFT}_1|\text{CFT}_2$), we can heuristically imagine folding along the interface to get a conformal boundary in the tensor product theory $\text{CFT}_1\otimes\overline{\text{CFT}}_2$, where $\overline{\text{CFT}}_2$ represents the theory $\text{CFT}_2$ with left- and right-movers interchanged (in practice we can accomplish this by a simple worldsheet time-reversal).  For our examples, the time reversal doesn't change the theory, but does act on individual states and operators.

Inverting this procedure, if we have a conformal boundary in the $\text{CFT}_1\otimes\overline{\text{CFT}}_2$ theory, then we can unfold it to obtain a conformal interface between $\text{CFT}_1$ and $\text{CFT}_2$.  Again, being somewhat schematic, if we can write
\be
|B\rangle\rangle=\sum_{\la_1,\widetilde{\la}_1,\la_2,\widetilde{\la}_2}B_{\la_1,\widetilde{\la}_1,\la_2,\widetilde{\la_2}}|\la_1,\widetilde{\la}_1\rangle_1\otimes|\la_2,\widetilde{\la}_2\rangle_2,
\ee
where the $\la_i$ and $\widetilde{\la}_i$ denote states in the left- and right-moving Hilbert spaces of $\text{CFT}_1$ and $\overline{\text{CFT}}_2$ respectively, and $B_{\la_1,\widetilde{\la}_1,\la_2,\widetilde{\la_2}}$ are real coefficients, then the corresponding interface has the form
\be
\mcO=\sum_{\la_1,\widetilde{\la}_1,\la_2,\widetilde{\la}_2}B_{\la_1,\widetilde{\la}_1,\la_2,\widetilde{\la_2}}|\la_1,\widetilde{\la}_1\rangle_1\langle\widetilde{\la}_2,\la_2|_2.
\ee
$\mcO$ maps operators in theory 2 to operators in theory 1.  Since we had a conformal boundary state satisfying
\be
0=\lp L_n^{tot}-\widetilde{L}_{-n}^{tot}\rp|B\rangle\rangle=\lp L_n^1+L_n^2-\widetilde{L}_{-n}^1-\widetilde{L}_{-n}^2\rp|B\rangle\rangle,
\ee
the interface will satisfy
\be
\lp L_n^1-\widetilde{L}_{-n}^1\rp\mcO=\mcO\lp L_n^2-\widetilde{L}_{-n}^2\rp,
\ee
and so it is called a conformal interface.  We will generically use $\mcO$ as the symbol for an interface, but when we want to stress that we have the same theory on either side of the interface (which can also be called a defect), we will sometimes use $\mcI$.

Explicitly, for the circle theory, by considering how time reversal acts on the mode expansion (\ref{eq:UntwistedModeExpansion}), one checks that the unfolding map acts as
\be
|N,M\rangle_2\rightarrow\langle -N,M|_2,\quad a_n^{2\,\dagger}\rightarrow -\wta_n^2,\quad\wta_n^{2\,\dagger}\rightarrow -a_n^2.
\ee

\subsection{$S^1|S^1$ interfaces}

We will first implement the unfolding procedure for the case of the $(S^1)^2$ boundary states we have constructed, following~\cite{bachas07}.  We must make some preliminary comments on notation.  An example of a term which appears in the expansions of our boundary states is
\be
e^{S_{12}^{(+)}a_n^{1\,\dagger}\wta_n^{2\,\dagger}}|-k_2N,k_1M\rangle_1\otimes|k_1N,k_2M\rangle_2.
\ee
If we are being completely explicit, then this gets sent, under the unfolding map, to
\be
\sum_{k=0}^\infty\frac{(-S_{12}^{(+)})^k}{k!}(a_n^{1\,\dagger})^k|-k_2N,k_1M\rangle_1\langle-k_1N,k_2M|_2(a_n^2)^k.
\ee
However, to save us from having to break every exponential up into explicit sums, we will abbreviate this expression as
\be
e^{-S_{12}^{(+)}a_n^{1\,\dagger}a_n^2}|-k_2N,k_1M\rangle_1\langle-k_1N,k_2M|_2.
\ee

With this understanding, we can write out the interfaces corresponding to the $B^{(+)}$ and $B^{(-)}$ classes of boundary states
\bea
\mcO^{(+)}_{k_1,k_2}(\al;\beta) &=& \sqrt{\frac{k_1^2R_1^2+k_2^2R_2^2}{2R_1R_2}}\lp\prod_{n=1}^\infty e^{S^{(+)}_{11}a_n^{1\,\dagger}\wta_n^{1\,\dagger}-S^{(+)}_{12}a_n^{1\,\dagger}a_n^2-S^{(+)}_{21}\wta_n^2\wta_n^{1\,\dagger}+S^{(+)}_{22}\wta_n^2a_n^2}\rp\non\\
&& \quad\times\sum_{N,M\in\Z}e^{-iM\al-iN\beta}|-k_2N,k_1M\rangle_1\langle -k_1N,k_2M|_2,\\
\mcO^{(-)}_{k_1,k_2}(\al;\beta) &=& \sqrt{\frac{k_1^2+4k_2^2R_1^2R_2^2}{4R_1R_2}}\lp\prod_{n=1}^\infty e^{S^{(-)}_{11}a_n^{1\,\dagger}\wta_n^{1\,\dagger}-S^{(-)}_{12}a_n^{1\,\dagger}a_n^2-S^{(-)}_{21}\wta_n^2\wta_n^{1\,\dagger}+S^{(-)}_{22}\wta_n^2a_n^2}\rp\non\\
&& \quad\times\sum_{N,M\in\Z}e^{-iM\al-iN\beta}|k_1M,-k_2N\rangle_1\langle -k_1N,k_2M|_2.
\eea

In the next section we will compute the fusion algebra of these conformal interfaces, but for now we would like to point out two particular classes.  Totally reflective interfaces satisfy
\be
\lp L_n^1-\widetilde{L}_{-n}^1\rp\mcO=0,\qquad\mcO\lp L_n^2-\widetilde{L}_{-n}^2\rp=0,
\ee
i.e.\ $\mcO$ is really a factorized product of boundary states, $\mcO\sim|B_1\rangle\rangle\langle\langle B_2|$.  For the class of interfaces above, these are simply the factorized cases with either $k_1=0$ or $k_2=0$.  For many purposes it will be useful to treat these cases separately from the non-factorized interfaces, so for the factorized interfaces we will explicitly write the factorized form, e.g.\
\be
|N(\al)\rangle\rangle_1\langle\langle N(\beta)|_2,\quad|N(\al)\rangle\rangle_1\langle\langle D(\beta)|_2,\quad|D(\al)\rangle\rangle_1\langle\langle N(\beta)|_2,\quad|D(\al)\rangle\rangle_1\langle\langle D(\beta)|_2.
\ee
Then if we write $\mcO^{(\eta)}_{k_1,k_2}(\al;\beta)$, we will be assuming that $k_1$ and $k_2$ are both nonzero.

The second class of interfaces of particular interest are the totally transmissive, or topological interfaces (see \cite{Fuchs:2015ska} for a complementary approach to the topological case), which satisfy
\be
L_n^1\mcO=\mcO L_n^2,\qquad\widetilde{L}_n^1\mcO=\mcO\widetilde{L}_n^2.
\ee
It turns out that this is the case precisely when $S^{(\pm)}$ is purely off-diagonal~\cite{bachas07}, i.e.\ when $\vartheta=\pm\frac{\pi}{4}$, for $\mcO^{(+)}$ interfaces, or $\theta=\pm\frac{\pi}{4}$ for $\mcO^{(-)}$ interfaces. In this case we have either
\be
\frac{k_1}{k_2}=\e\frac{R_2}{R_1},\qquad\mathrm{or}\qquad\frac{k_1}{k_2}=2\e R_1R_2,
\ee
for $\mcO^{(+)}$ or $\mcO^{(-)}$ interfaces respectively, and where $\e=\pm 1$.  Note that this can only happen at specific radii, i.e.\ when $R_2$ is a rational multiple of either $R_1$ or of $1/R_1$.  In these cases the given interfaces are topological.  As we will see when we compute the fusion algebras, global symmetries are implemented by invertible topological interfaces, and in general topological interfaces have nice properties under fusion.

\subsection{$S^1|(S^1/\Z_2)$ interfaces}

For generic boundary states, the interfaces are essentially just given by symmetrizing the circle interfaces,
\be
\mcO^{(\eta)\,SO}_{k_1,k_2}(\al;\beta)=\frac{1}{\sqrt{2}}\mcO^{(\eta)}_{k_1,k_2}(\al;\beta)+\frac{1}{\sqrt{2}}\mcO^{(\eta)}_{k_1,-k_2}(\al;-\beta).
\ee
The $SO$ superscript indicates that we have an $S^1$ theory on the left and an orbifold theory on the right of the interface.  We will use $OS$ and $OO$ superscripts in the same way, and if a superscript is omitted then it implicitly means we have $SS$.

The non-generic states only occur when either $k_1$ or $k_2$ vanishes, so these are only factorized interfaces,
\be
|N(\al)\rangle\rangle_1\langle\langle N_O(\beta)|_2,\quad |N(\al)\rangle\rangle_1\langle\langle N_O(\beta_0,\e)\rangle\rangle_2,\quad |N(\al)\rangle\rangle_1\langle\langle D_O(\beta)|_2,\quad |N(\al)\rangle\rangle_1\langle\langle D_O(\beta_0,\e)|_2,\non
\ee
\be
|D(\al)\rangle\rangle_1\langle\langle N_O(\beta)|_2,\quad |D(\al)\rangle\rangle_1\langle\langle N_O(\beta_0,\e)\rangle\rangle_2,\quad |D(\al)\rangle\rangle_1\langle\langle D_O(\beta)|_2,\quad |D(\al)\rangle\rangle_1\langle\langle D_O(\beta_0,\e)|_2.
\ee

The topological defects again occur when the angle $\vartheta$ or $\theta$ equals $\pi/4$, which implies that
\be
\label{eq:TopRequirement}
\frac{k_1}{k_2}=\frac{R_2}{R_1},\qquad\mathrm{or}\qquad\frac{k_1}{k_2}=2R_1R_2.
\ee
Note that these are all generic states, and don't involve any twisted sector pieces.  This means that for this case the topological defects all act as zero on the twisted sector part of the $S^1/\Z_2$ Hilbert space.  The only interfaces we have found that don't involve a projection operator onto the untwisted sector are the factorized, totally reflective interfaces.

Of course, interfaces of $S^1/\Z_2|S^1$ are completely analogous, with the obvious adjustments.  In particular, the generic states are now
\be
\mcO^{(\eta)\,OS}_{k_1,k_2}(\al;\beta)=\frac{1}{\sqrt{2}}\mcO^{(\eta)}_{k_1,k_2}(\al;\beta)+\frac{1}{\sqrt{2}}\mcO^{(\eta)}_{k_1,-k_2}(-\al;\beta).
\ee

\subsection{$(S^1/\Z_2)|(S^1/\Z_2)$ interfaces}

For conformal interfaces between two $S^1/\Z_2$ theories, we have the generic interfaces, which are simply linear combinations of $S^1$ interfaces, e.g.\
\be
\mcO^{(\eta)\,OO}_{k_1,k_2}(\al;\beta)=\hlf\mcO^{(\eta)}_{k_1,k_2}(\al;\beta)+\hlf\mcO^{(\eta)}_{k_1,-k_2}(-\al;\beta)+\hlf\mcO^{(\eta)}_{k_1,-k_2}(\al;-\beta)+\hlf\mcO^{(\eta)}_{k_1,k_2}(-\al;-\beta).
\ee
We also have the factorized, or totally reflective, interfaces which are related to factorized products of boundary states such as
\be
|N_O(\al_0,\e)\rangle\rangle_1\langle\langle D_O(\beta_0,\e')|_2,\qquad|N_O(\beta)\rangle\rangle_1\langle\langle N_O(0,\e)|_2,
\ee
etc.  Finally, in this case we have a third category of interfaces when $k_1$ and $k_2$ are both positive, and $\al$ and $\beta$ are both either $0$ or $\pi$.  These have the form
\be
\mcO^{(\eta)\,OO}_{k_1,k_2}(\al_0;\beta_0;\e)=\hlf\mcO^{(\eta)}_{k_1,k_2}(\al_0;\beta_0)+\hlf\mcO^{(\eta)}_{k_1,-k_2}(\al_0;\beta_0)+\e\mcO^{(\eta)\,OO}_{tw}.
\ee
In this expression, the final term is a linear combination of terms of the form
\begin{multline}
\hlf\ls\lp\prod_{r\in\Z+\hlf,r>0}e^{S^{(\eta)}_{11}a_r^{1\,t\,\dagger}\wta_r^{1\,t\,\dagger}-S^{(\eta)}_{12}a_r^{1\,t\,\dagger}a_r^{2\,t}-S^{(\eta)}_{21}\wta_r^{2\,t}\wta_r^{1\,t\,\dagger}+S^{(\eta)}_{22}\wta_r^{2\,t}a_r^{2\,t}}\rp\right.\\
\left.+\lp\prod_{r\in\Z+\hlf,r>0}e^{S^{(\eta)}_{11}a_r^{1\,t\,\dagger}\wta_r^{1\,t\,\dagger}+S^{(\eta)}_{12}a_r^{1\,t\,\dagger}a_r^{2\,t}+S^{(\eta)}_{21}\wta_r^{2\,t}\wta_r^{1\,t\,\dagger}+S^{(\eta)}_{22}\wta_r^{2\,t}a_r^{2\,t}}\rp\rs|\g_0\rangle_1\langle\d_0|_2,
\end{multline}
where $\g_0\in\{0,\pi R_1\}$, $\d_0\in\{0,\pi R_2\}$, and which linear combination we should take for given $\al_0,\beta_0$ and $\eta$ can be determined from the results enumerated in Table \ref{tab:vecs}.

None of these last classes are reflective, but they can be topological, under the same conditions (\ref{eq:TopRequirement}) as before.  In fact, these cases will play a prominent role as we identify the interesting defects in this theory in the next section.

As with the corresponding $(S^1/\Z_2)^2$ boundary states, we have threshold relations
\be
\label{eq:NDOOThreshold}
\lim_{\al\rr\al_0;\beta\rr\beta_0}\mcO^{(\eta)\,OO}_{k_1,k_2}(\al;\beta)=\mcO^{(\eta)\,OO}_{k_1,k_2}(\al_0;\beta_0;+)+\mcO^{(\eta)\,OO}_{k_1,k_2}(\al_0;\beta_0;-).
\ee

\section{Fusion of interfaces}
\label{sec:Fusion}

Part of what makes the structure of conformal interfaces so rich is that they can be fused together.  An interface $\mcO$ from $\text{CFT}_2$ to $\text{CFT}_1$ can be combined with an interface $\mcO'$ from $\text{CFT}_3$ to $\text{CFT}_2$ to obtain a new conformal interface $\mcO''=\mcO\circ\mcO'$ going from $\text{CFT}_3$ to $\text{CFT}_1$.  We define the fusion by regularizing the ordinary product,
\be
\label{eq:RegularizedFusion}
\mcO\circ\mcO'=\lim_{t\rr 0}e^{2\pi d/t}\mcO e^{-tH_2}\mcO',
\ee
where $H_2$ is the Hamiltonian for $\text{CFT}_2$, and $d$ is a constant determined uniquely by the requirement that we get a finite result.  If either defect is topological, then regularization is not necessary (i.e.\ $d=0$), and the fusion can be defined by the ordinary product of operators.  For example, if $\mcO'$ is topological, then
\be
\mcO\circ\mcO'=\lim_{t\rr 0}\mcO e^{-tH_2}\mcO'=\lim_{t\rr 0}\mcO\mcO'e^{-tH_3}=\mcO\mcO'.
\ee

Let us also note that if either interface is totally reflective, then the fusion is also totally reflective, while if both interfaces are topological, then the fusion is topological.

\subsection{Fusion of factorized interfaces}

In the case where both interfaces are factorized, then the fusion is very simple, and can be read off from the annulus diagram computations of section \ref{sec:BCFT1}.  In particular, the regularization procedure amounts to simply taking the coefficient of the leading term in the $\wtq$ expansion.  For $S^1$ boundary states, this is
\bea
\lp|B\rangle\rangle_1\langle\langle N(\al)|_2\rp\circ\lp|N(\al')\rangle\rangle_2\langle\langle B'|_3\rp &=& \left\{\begin{matrix} 2|B\rangle\rangle_1\langle\langle B'|_3, & \mathrm{if\ }|\al'-\al|=\pi,\\ |B\rangle\rangle_1\langle\langle B'|_3, & \mathrm{otherwise.}\end{matrix}\right.\\
\lp|B\rangle\rangle_1\langle\langle N(\al)|_2\rp\circ\lp|D(\beta')\rangle\rangle_2\langle\langle B'|_3\rp &=& |B\rangle\rangle_1\langle\langle B'|_3,\\
\lp|B\rangle\rangle_1\langle\langle D(\beta)|_2\rp\circ\lp|N(\al')\rangle\rangle_2\langle\langle B'|_3\rp &=& |B\rangle\rangle_1\langle\langle B'|_3,\\
\lp|B\rangle\rangle_1\langle\langle D(\beta)|_2\rp\circ\lp|D(\beta')\rangle\rangle_2\langle\langle B'|_3\rp &=& \left\{\begin{matrix} 2|B\rangle\rangle_1\langle\langle B'|_3, & \mathrm{if\ }|\beta'-\beta|=\pi,\\ |B\rangle\rangle_1\langle\langle B'|_3, & \mathrm{otherwise.}\end{matrix}\right.
\eea
Here $|B\rangle\rangle_1$ and $\langle\langle B'|_3$ can be any boundary states, either in the $S^1$ or $S^1/\Z_2$ theory.  Note that the fusion coefficients are not continuous as we vary the parameters $\al$ and $\al'$ or $\beta$ and $\beta'$.  This discontinuity is an artifact of the regularization procedure.  By this we mean that if we defined a $t$-dependent product
\be
\mcO\diamond_t\mcO'=\mcO e^{-tH_2}\mcO',
\ee
then there is no discontinuity as we vary parameters.  For the factorized states, the effect of the regularization procedure (\ref{eq:RegularizedFusion}) is to isolate the contribution of the ground state in the open-string Hilbert space.  The discontinuity arises when there is a cross-over of the open-string ground state and the first excited state in the annulus diagram.  The ground state then becomes degenerate, and this results in a jump in the fusion coefficients.  Since topological interfaces do not require regularization, we will see that their fusion is continuous.

If the inner states are $S^1/\Z_2$ boundary states, then using the results of \ref{app:OrbifoldAnnulus} (we omit results that are related by exchanging left and right theories)
\bea
\lp|B\rangle\rangle_1\langle\langle |N_O(\al)|_2\rp\circ\lp N_O(\al')\rangle\rangle_2\langle\langle B'|_3\rp &=& |B\rangle\rangle_1\langle\langle B'|_3,\\
\lp|B\rangle\rangle_1\langle\langle |N_O(\al)|_2\rp\circ\lp N_O(\al_0',\e')\rangle\rangle_2\langle\langle B'|_3\rp &=& |B\rangle\rangle_1\langle\langle B'|_3,\\
\lp|B\rangle\rangle_1\langle\langle |N_O(\al)|_2\rp\circ\lp D_O(\beta')\rangle\rangle_2\langle\langle B'|_3\rp &=& 2|B\rangle\rangle_1\langle\langle B'|_3,\\
\lp|B\rangle\rangle_1\langle\langle |N_O(\al)|_2\rp\circ\lp D_O(\beta_0',\e')\rangle\rangle_2\langle\langle B'|_3\rp &=& |B\rangle\rangle_1\langle\langle B'|_3,\\
\label{eq:NNFusion}
\lp|B\rangle\rangle_1\langle\langle |N_O(\al_0,\e)|_2\rp\circ\lp N_O(\al_0',\e')\rangle\rangle_2\langle\langle B'|_3\rp &=& \left\{\begin{matrix}2|B\rangle\rangle_1\langle\langle B'|_3, & \mathrm{if\ }\al_0'=\al_0,\ \e'=-\e,\ R_2=1,\\ |B\rangle\rangle_1\langle\langle B'|_3, & \mathrm{otherwise,}\end{matrix}\right.\non\\
\\
\lp|B\rangle\rangle_1\langle\langle |N_O(\al_0,\e)|_2\rp\circ\lp D_O(\beta')\rangle\rangle_2\langle\langle B'|_3\rp &=& |B\rangle\rangle_1\langle\langle B'|_3,\\
\lp|B\rangle\rangle_1\langle\langle |N_O(\al_0,\e)|_2\rp\circ\lp D_O(\beta_0',\e')\rangle\rangle_2\langle\langle B'|_3\rp &=& |B\rangle\rangle_1\langle\langle B'|_3,\\
\lp|B\rangle\rangle_1\langle\langle |D_O(\beta)|_2\rp\circ\lp D_O(\beta')\rangle\rangle_2\langle\langle B'|_3\rp &=& |B\rangle\rangle_1\langle\langle B'|_3,\\
\lp|B\rangle\rangle_1\langle\langle |D_O(\beta)|_2\rp\circ\lp D_O(\beta_0',\e')\rangle\rangle_2\langle\langle B'|_3\rp &=& |B\rangle\rangle_1\langle\langle B'|_3,\\
\label{eq:DDFusion}
\lp|B\rangle\rangle_1\langle\langle |D_O(\beta_0,\e)|_2\rp\circ\lp D_O(\beta_0',\e')\rangle\rangle_2\langle\langle B'|_3\rp &=& \left\{\begin{matrix}2|B\rangle\rangle_1\langle\langle B'|_3, & \mathrm{if\ }\al_0'=\al_0,\ \e'=-\e,\ R_2=\hlf,\\ |B\rangle\rangle_1\langle\langle B'|_3, & \mathrm{otherwise.}\end{matrix}\right.\non\\
\eea
Note that if we restirct to $R_2>1/\sqrt{2}$, then the $R_2=1/2$ case in (\ref{eq:DDFusion}) doesn't occur, and regardless it is the T-dual of the $R_2=1$ case in (\ref{eq:NNFusion}).

\subsection{Fusion of one non-factorized and one factorized interface}

This computation is somewhat more involved, and can mostly be obtained as a special case of the fully non-factorized computation.  In particular, the oscillator contribution is obtained by specialization at the end of Appendix \ref{appsubsec:OscillatorFusion}, and the zero mode contribution is worked out in Appendix \ref{appsubsec:ZMFusionSpecial}.

The result for $S^1|S^1$ interfaces is~\cite{bachas07,Bachas:2012bj}
\bea
\mcO^{(+)}_{k_1,k_2}(\al;\beta)\circ\lp|N(\al')\rangle\rangle_2\langle\langle B'|_3\rp &=& \sum_{j=0}^{|k_1|-1}|N(\frac{\al+k_2\al'+2\pi j}{k_1})\rangle\rangle_1\langle\langle B'|_3,\\
\mcO^{(+)}_{k_1,k_2}(\al;\beta)\circ\lp|D(\beta')\rangle\rangle_2\langle\langle B'|_3\rp &=& \sum_{j=0}^{|k_2|-1}|D(\frac{-\beta+k_1\beta'+2\pi j}{k_2})\rangle\rangle_1\langle\langle B'|_3,\\
\mcO^{(-)}_{k_1,k_2}(\al;\beta)\circ\lp|N(\al')\rangle\rangle_2\langle\langle B'|_3\rp &=& \sum_{j=0}^{|k_1|-1}|D(\frac{\al+k_2\al'+2\pi j}{k_1})\rangle\rangle_1\langle\langle B'|_3,\\
\mcO^{(-)}_{k_1,k_2}(\al;\beta)\circ\lp|D(\beta')\rangle\rangle_2\langle\langle B'|_3\rp &=& \sum_{j=0}^{|k_2|-1}|N(\frac{-\beta+k_1\beta'+2\pi j}{k_2})\rangle\rangle_1\langle\langle B'|_3.
\eea

When the non-factorized interface is of $OS$ type, the modification is simple,
\bea
\mcO^{(+)\,OS}_{k_1,k_2}(\al;\beta)\circ\lp|N(\al')\rangle\rangle_2\langle\langle B'|_3\rp &=& \sum_{j=0}^{|k_1|-1}|N_O(\frac{\al+k_2\al'+2\pi j}{k_1})\rangle\rangle_1\langle\langle B'|_3,\\
\mcO^{(+)\,OS}_{k_1,k_2}(\al;\beta)\circ\lp|D(\beta')\rangle\rangle_2\langle\langle B'|_3\rp &=& \sum_{j=0}^{|k_2|-1}|D_O(\frac{-\beta+k_1\beta'+2\pi j}{k_2})\rangle\rangle_1\langle\langle B'|_3,\\
\mcO^{(-)\,OS}_{k_1,k_2}(\al;\beta)\circ\lp|N(\al')\rangle\rangle_2\langle\langle B'|_3\rp &=& \sum_{j=0}^{|k_1|-1}|D_O(\frac{\al+k_2\al'+2\pi j}{k_1})\rangle\rangle_1\langle\langle B'|_3,\\
\mcO^{(-)\,OS}_{k_1,k_2}(\al;\beta)\circ\lp|D(\beta')\rangle\rangle_2\langle\langle B'|_3\rp &=& \sum_{j=0}^{|k_2|-1}|N_O(\frac{-\beta+k_1\beta'+2\pi j}{k_2})\rangle\rangle_1\langle\langle B'|_3,
\eea
with the understanding that we must use the threshold relation (\ref{eq:O1Threshold}) when any of the arguments on the right hand side coincide with either $0$ or $\pi$.

Similarly, if the non-factorized interface is $SO$ and the boundary state is of orbifold type, we get
\bea
\mcO^{(+)\,SO}_{k_1,k_2}(\al;\beta)\circ\lp|N_O(\al')\rangle\rangle_2\langle\langle B'|_3\rp &=& \sum_{j=0}^{|k_1|-1}\lp|N(\frac{\al+k_2\al'+2\pi j}{k_1})\rangle\rangle_1\langle\langle B'|_3\right.\non\\
&& \qquad\left. +|N(\frac{\al-k_2\al'+2\pi j}{k_1})\rangle\rangle_1\langle\langle B'|_3\rp,\\
\mcO^{(+)\,SO}_{k_1,k_2}(\al;\beta)\circ\lp|D_O(\beta')\rangle\rangle_2\langle\langle B'|_3\rp &=& \sum_{j=0}^{|k_2|-1}\lp|D(\frac{-\beta+k_1\beta'+2\pi j}{k_2})\rangle\rangle_1\langle\langle B'|_3\right.\non\\
&& \qquad\left. +|D(\frac{-\beta-k_1\beta'+2\pi j}{k_2})\rangle\rangle_1\langle\langle B'|_3\rp,\\
\mcO^{(-)\,SO}_{k_1,k_2}(\al;\beta)\circ\lp|N_O(\al')\rangle\rangle_2\langle\langle B'|_3\rp &=& \sum_{j=0}^{|k_1|-1}\lp|D(\frac{\al+k_2\al'+2\pi j}{k_1})\rangle\rangle_1\langle\langle B'|_3\right.\non\\
&& \qquad\left. +|D(\frac{\al-k_2\al'+2\pi j}{k_1})\rangle\rangle_1\langle\langle B'|_3\rp,\\
\mcO^{(-)\,SO}_{k_1,k_2}(\al;\beta)\circ\lp|D_O(\beta')\rangle\rangle_2\langle\langle B'|_3\rp &=& \sum_{j=0}^{|k_2|-1}\lp|N(\frac{-\beta+k_1\beta'+2\pi j}{k_2})\rangle\rangle_1\langle\langle B'|_3\right.\non\\
&& \qquad\left. +|N(\frac{-\beta-k_1\beta'+2\pi j}{k_2})\rangle\rangle_1\langle\langle B'|_3\rp,
\eea
or
\bea
\mcO^{(+)\,SO}_{k_1,k_2}(\al;\beta)\circ\lp|N_O(\al_0',\e')\rangle\rangle_2\langle\langle B'|_3\rp &=& \sum_{j=0}^{|k_1|-1}|N(\frac{\al+k_2\al_0'+2\pi j}{k_1})\rangle\rangle_1\langle\langle B'|_3,\\
\mcO^{(+)\,SO}_{k_1,k_2}(\al;\beta)\circ\lp|D_O(\beta_0',\e')\rangle\rangle_2\langle\langle B'|_3\rp &=& \sum_{j=0}^{|k_2|-1}|D(\frac{-\beta+k_1\beta_0'+2\pi j}{k_2})\rangle\rangle_1\langle\langle B'|_3,\\
\mcO^{(-)\,SO}_{k_1,k_2}(\al;\beta)\circ\lp|N_O(\al_0',\e')\rangle\rangle_2\langle\langle B'|_3\rp &=& \sum_{j=0}^{|k_1|-1}|D(\frac{\al+k_2\al_0'+2\pi j}{k_1})\rangle\rangle_1\langle\langle B'|_3,\\
\mcO^{(-)\,SO}_{k_1,k_2}(\al;\beta)\circ\lp|D_O(\beta_0',\e')\rangle\rangle_2\langle\langle B'|_3\rp &=& \sum_{j=0}^{|k_2|-1}|N(\frac{-\beta+k_1\beta_0'+2\pi j}{k_2})\rangle\rangle_1\langle\langle B'|_3,
\eea

Finally, we have fusion involving $OO$ interfaces.  When both the non-factorized and factorized interfaces have generic parameters, we find
\bea
\mcO^{(+)\,OO}_{k_1,k_2}(\al;\beta)\circ\lp|N_O(\al')\rangle\rangle_2\langle\langle B'|_3\rp &=& \sum_{j=0}^{|k_1|-1}\lp|N_O(\frac{\al+k_2\al'+2\pi j}{k_1})\rangle\rangle_1\langle\langle B'|_3\right.\non\\
&& \qquad\left. +|N_O(\frac{\al-k_2\al'+2\pi j}{k_1})\rangle\rangle_1\langle\langle B'|_3\rp,\\
\mcO^{(+)\,OO}_{k_1,k_2}(\al;\beta)\circ\lp|D_O(\beta')\rangle\rangle_2\langle\langle B'|_3\rp &=& \sum_{j=0}^{|k_2|-1}\lp|D_O(\frac{-\beta+k_1\beta'+2\pi j}{k_2})\rangle\rangle_1\langle\langle B'|_3\right.\non\\
&& \qquad\left. +|D_O(\frac{-\beta-k_1\beta'+2\pi j}{k_2})\rangle\rangle_1\langle\langle B'|_3\rp,\\
\mcO^{(-)\,OO}_{k_1,k_2}(\al;\beta)\circ\lp|N_O(\al')\rangle\rangle_2\langle\langle B'|_3\rp &=& \sum_{j=0}^{|k_1|-1}\lp|D_O(\frac{\al+k_2\al'+2\pi j}{k_1})\rangle\rangle_1\langle\langle B'|_3\right.\non\\
&& \qquad\left. +|D_O(\frac{\al-k_2\al'+2\pi j}{k_1})\rangle\rangle_1\langle\langle B'|_3\rp,\\
\mcO^{(-)\,OO}_{k_1,k_2}(\al;\beta)\circ\lp|D_O(\beta')\rangle\rangle_2\langle\langle B'|_3\rp &=& \sum_{j=0}^{|k_2|-1}\lp|N_O(\frac{-\beta+k_1\beta'+2\pi j}{k_2})\rangle\rangle_1\langle\langle B'|_3\right.\non\\
&& \qquad\left. +|N_O(\frac{-\beta-k_1\beta'+2\pi j}{k_2})\rangle\rangle_1\langle\langle B'|_3\rp,
\eea
making use of threshold relations as appropriate.  When only one of the two interfaces is non-generic, we get only the first of the two terms on the right hand side.

If both the non-factorized and factorized interfaces are non-generic, then the fusion must be calculated on a case by case basis.  Rather than enumerate all possibilities, we will focus on the case where the non-factorized interface is $\mcO^{(+)\,OO}_{1,1}(\al_0;\beta_0;\e)$, because these are of the most interest below.  We have
\bea
\mcO^{(+)\,OO}_{1,1}(\al_0;\beta_0;\e)\circ\lp|N_O(\al_0',\e')\rangle\rangle_2\langle\langle B'|_3\rp &=& |N_O(\al_0+\al_0',\lp -1\rp^{\frac{\beta_0\al_0'}{\pi^2}}\e\e')\rangle\rangle_1\langle\langle B'|_3,\\
\mcO^{(+)\,OO}_{1,1}(\al_0;\beta_0;\e)\circ\lp|D_O(\beta_0',\e')\rangle\rangle_2\langle\langle B'|_3\rp &=& |D_O(\beta_0+\beta_0',\lp -1\rp^{\frac{\al_0(\beta_0+\beta_0')}{\pi^2}}\e\e')\rangle\rangle_1\langle\langle B'|_3.\non\\
\eea
In particular, we note that $\mcI^{(+)\,OO}_{1,1}(0;0;+)$ acts as the identity in these fusion products, mapping each factorized interface to itself.

\subsection{Fusion of non-factorized $S^1|S^1$ interfaces}
\label{subsec:S1Fusion}

In Appendix \ref{subapp:S1Fusion}, we review the computation of the fusion for the $S^1|S^1$ interfaces, as first done by~\cite{bachas07,Bachas:2012bj}.  Here we quote the results,
\be
\label{eq:S1Fusion}
\mcO^{(\eta)}_{k_1,k_2}(\al;\beta)\circ\mcO^{(\eta')}_{k_1',k_2'}(\al';\beta')=\sum_{j=0}^{G_1^{(\eta')}-1}\sum_{j'=0}^{G_2^{(\eta')}-1}\mcO^{(\eta\eta')}_{K_1^{(\eta')},K_2^{(\eta')}}(\al^{(\eta')}+\frac{2\pi j'}{G_2^{(\eta')}};\beta^{(\eta')}+\frac{2\pi j}{G_1^{(\eta')}}),
\ee
where
\be
\label{eq:GDefs}
G_1^{(+)}=\gcd(k_2,k_1'),\qquad G_2^{(+)}=\gcd(k_1,k_2'),\qquad G_1^{(-)}=\gcd(k_1,k_1'),\qquad G_2^{(-)}=\gcd(k_2,k_2'),
\ee
\be
\label{eq:PlusDefs}
K_1^{(+)}=\frac{k_1k_1'}{G_1^{(+)}G_2^{(+)}},\quad K_2^{(+)}=\frac{k_2k_2'}{G_1^{(+)}G_2^{(+)}},\qquad\al^{(+)}=\frac{k_1'\al+k_2\al'}{G_1^{(+)}G_2^{(+)}},\quad\beta^{(+)}=\frac{k_2'\beta+k_1\beta'}{G_1^{(+)}G_2^{(+)}},
\ee
and
\be
\label{eq:MinusDefs}
K_1^{(-)}=\frac{k_2k_1'}{G_1^{(-)}G_2^{(-)}},\quad K_2^{(-)}=\frac{k_1k_2'}{G_1^{(-)}G_2^{(-)}},\qquad\al^{(-)}=\frac{k_1\al'-k_1'\beta}{G_1^{(-)}G_2^{(-)}},\quad\beta^{(-)}=\frac{k_2\beta'-k_2'\al}{G_1^{(-)}G_2^{(-)}}.
\ee
One can confirm that the products of two topological interfaces are topological.  For example, if
\be
\frac{k_1}{k_2}=\zeta\frac{R_2}{R_1},\qquad\frac{k_1'}{k_2'}=2\zeta'R_2R_3,
\ee
where $\zeta$ and $\zeta'$ are $\pm 1$, then
\be
\frac{K_1^{(-)}}{K_2^{(-)}}=\frac{k_2k_1'}{k_1k_2'}=2\zeta\zeta'R_1R_3,
\ee
so the results are topological.

Let's now restrict to interfaces from one theory to itself, with $R_1=R_2=R$.  In the language of~\cite{Bachas:2012bj} these interfaces are called defects, and the collection of such defects are closed under multiplication and addition, so they form a semigroup, or monoid, called the {\it{defect monoid}}.  In our case the topological $\eta=+$ defects correspond to $k_1=1$, $k_2=\zeta=\pm 1$.  These defects are not only topological, but are also invertible, and generate the generic global symmetry $\U(1)^2\rtimes\Z_2$ for the free boson at radius $R$, and their fusions simply realize this symmetry group,
\be
\mcI^{(+)}_{1,\zeta}(\al;\beta)\circ\mcI^{(+)}_{1,\zeta'}(\al';\beta')=\mcI^{ND}_{1,\zeta\zeta'}(\al+\zeta\al';\beta'+\zeta'\beta).
\ee
Note in particular that $\mcI^{(+)}_{1,1}(0;0)$ is the identity defect,
\be
\mcI^{(+)}_{1,1}(0;0)\circ\mcO^{(\eta)}_{k_1,k_2}(\al;\beta)=\mcO^{(\eta)}_{k_1,k_2}(\al;\beta),\quad\mcO^{(\eta)}_{k_1,k_2}(\al;\beta)\circ\mcI^{(+)}_{1,1}(0;0)=\mcO^{(\eta)}_{k_1,k_2}(\al;\beta).
\ee
Another perspective on the identity interface is that if we write it out explicitly, we see that it essentially represents the insertion of a complete set of normalized states,
\be
\mcI^{(+)}_{1,1}(0;0)=\sum_{|\psi\rangle\in\mathcal{H}}|\psi\rangle\langle\psi|.
\ee
The interfaces $\mcI^{(+)}_{1,1}(\al;\beta)$ implement translation or dual translation, and the interface $\mcI^{(+)}_{1,-1}(0;0)$ implements the reflection symmetry.

The $\eta=-$ interfaces can only be topological at special values of $R$, when $R^2$ is rational, and they are only invertible at the self-dual radius $R=1/\sqrt{2}$.  At the self-dual radius however, the $\mcI^{(-)}_{1,\pm 1}(\al;\beta)$ do enhance our set of global symmetry interfaces,
\bea
\mcI^{(+)}_{1,\zeta}(\al;\beta)\circ\mcI^{(-)}_{1,\zeta'}(\al';\beta') &=& \mcI^{(-)}_{1,\zeta\zeta'}(\zeta\al'-\zeta\beta;\beta'-\zeta\zeta'\al),\\
\mcI^{(-)}_{1,\zeta}(\al;\beta)\circ\mcI^{(+)}_{1,\zeta'}(\al';\beta') &=& \mcI^{(-)}_{1,\zeta\zeta'}(\al+\zeta\al';\beta'+\zeta'\beta),\\
\mcI^{(-)}_{1,\zeta}(\al;\beta)\circ\mcI^{(-)}_{1,\zeta'}(\al';\beta') &=& \mcI^{(+)}_{1,\zeta\zeta'}(\zeta\al'-\zeta\beta;\beta'-\zeta\zeta'\al).
\eea
These fusions hold at any radius (with $\mcI$ replaced by $\mcO$), but they are only topological at the self-dual radius.  Note that although this has expanded the implemented global symmetries to $(\U(1)\rtimes\Z_2)^2$, but this is still significantly less than the expected enhanced symmetry group of $\SU(2)^2/\Z_2$.  This indicates that we do not have a full enumeration of interfaces in general.

Another important interface is the radius-changing interface, $\mcO^{(+)}_{1,1}(0;0)$.  If $R_2=R_1$, then it reduces to the identity interface discussed above, but for $R_2\ne R_1$ it is the (non-topological) interface which implements the changing of the radius from $R_2$ to $R_1$.

Finally, we would like to point out one more set of interesting interfaces.  Consider the case when $R_1=R_2/m$, for some positive integer $m$.  Then $\mcO^{(+)}_{m,\pm 1}(\al;\beta)$ are topological.  Note that the $R_1$ theory can be obtained from the $R_2$ theory by performing a $\Z_m$ orbifold, translating by $2\pi R_2/m$ around the circle.  We claim that $\mcO^{(+)}_{m,1}(0;0)$ (the other topological interfaces mentioned above can be obtained by acting on this one with a global symmetry interface) implements the mapping between invariant states in the $R_2$ theory and untwisted sector states in the orbifold theory.  Indeed,
\be
\mcO^{(+)}_{m,1}(0;0)|N,M\rangle_2=\left\{\begin{matrix}\sqrt{m}|\frac{N}{m},mM\rangle_1, & \mathrm{if\ }N\equiv 0\ (\mathrm{mod}\ m),\\ 0, & \mathrm{otherwise}.\end{matrix}\right.
\ee
Conversely, if $R_1=mR_2$, then there is a topological interface $\mcO^{(+)}_{1,m}(0;0)$ which does the same for the quantum symmetry orbifold.  The fusion of these two ``orbifold interfaces'' is
\be
\mcO^{(+)}_{1,m}(0;0)\circ\mcO^{(+)}_{m,1}(0;0)=\sum_{j=0}^{m-1}\mcI^{(+)}_{1,1}(0;\frac{2\pi j}{m})=m\Pi_{\Z_m},
\ee
where for any group $G$
\be
\Pi_G=\frac{1}{|G|}\sum_{g\in G}\rho(g),
\ee
is the projector onto $G$-invariant states.  In the case at hand, it is the $\Z_m$ group of translation by $2\pi R_2/m$.  Similarly,
\be
\mcO^{(+)}_{m,1}(0;0)\circ\mcO^{(+)}_{1,m}(0;0)=\sum_{j'=0}^{m-1}\mcO^{(+)}_{1,1}(\frac{2\pi j'}{m};0)=m\Pi_{\Z_m'},
\ee
projection onto the group of $\Z_m$ dual translations.

\subsection{Fusion involving orbifold interfaces}

Using the results for fusion of $S^1|S^1$ interfaces, we can easily compute the fusion of any two non-factorized interfaces as long as at least one of them is generic.  We can summarize the results as follows
\bea
\mcO\circ\mcO^{\prime\,SO} &=& \sum\sum\widetilde{\mcO}^{SO},\\
\mcO^{OS}\circ\mcO' &=& \sum\sum\widetilde{\mcO}^{OS},\\
\mcO^{SO}\circ\mcO^{\prime\,OS} &=& \sum\sum\lp\widetilde{\mcO}+\widehat{\mcO}\rp,\\
\mcO^{OS}\circ\mcO^{\prime\,SO} &=& \sum\sum\widetilde{\mcO}^{OO},\\
\mcO^{SO}\circ\mcO^{\prime\,OO} &=& \sum\sum\lp\widetilde{\mcO}^{SO}+\widehat{\mcO}^{SO}\rp,\\
\mcO^{OO}\circ\mcO^{\prime\,OS} &=& \sum\sum\lp\widetilde{\mcO}^{OS}+\widehat{\mcO}^{OS}\rp,\\
\mcO^{OO}\circ\mcO^{\prime\,OO} &=& \sum\sum\lp\widetilde{\mcO}^{OO}+\widehat{\mcO}^{OO}\rp,\\
\mcO^{SO}\circ\mcO^{\prime\,OO}_\e &=& \sum\sum\widetilde{\mcO}^{SO},\\
\mcO^{OO}_\e\circ\mcO^{\prime\,OS} &=& \sum\sum\widetilde{\mcO}^{OS},\\
\mcO^{OO}\circ\mcO^{\prime\,OO}_\e &=& \sum\sum\widetilde{\mcO}^{OO},\\
\mcO^{OO}_\e\circ\mcO^{\prime\,OO} &=& \sum\sum\widetilde{\mcO}^{OO},
\eea
where we have used shorthand
\be
\mcO^{XY}=\mcO^{(\eta)\,XY}_{k_1,k_2}(\al;\beta),\qquad\mcO^{\prime\,XY}=\mcO^{(\eta')\,XY}_{k_1',k_2'}(\al';\beta'),\non
\ee
\be
\mcO^{OO}_\e=\mcO^{(\eta)\,OO}_{k_1,k_2}(\al;\beta;\e),\qquad\mcO^{\prime\,OO}_\e=\mcO^{(\eta')\,OO}_{k_1',k_2'}(\al';\beta';\e),\non
\ee
\be
\sum\sum\sim\sum_{j=0}^{G_1^{(\eta')}-1}\sum_{j'=0}^{G_2^{(\eta')}-1},\qquad\widetilde{\mcO}^{XY}=\mcO^{(\eta\eta')\,XY}_{K_1^{(\eta')},K_2^{(\eta')}}(\al^{(\eta')}+\frac{2\pi j'}{G_2^{(\eta')}};\beta^{(\eta')}+\frac{2\pi j}{G_1^{(\eta')}}),\non
\ee
\be
\widehat{\mcO}^{XY}=\mcO^{(\eta\eta')\,XY}_{K_1^{(\eta')},K_2^{(\eta')}}(\widehat{\al}^{(\eta')}+\frac{2\pi j'}{G_2^{(\eta')}};\widehat{\beta}^{(\eta')}+\frac{2\pi j}{G_1^{(\eta')}}),
\ee
and
\be
\widehat{\al}^{(+)}=\frac{k_1'\al-k_2\al'}{G_1^{(+)}G_2^{(+)}},\quad\widehat{\beta}^{(+)}=\frac{k_2'\beta-k_1\beta'}{G_1^{(+)}G_2^{(+)}},\qquad\widehat{\al}^{(-)}=\frac{-k_1\al'-k_1'\beta}{G_1^{(-)}G_2^{(-)}},\quad\widehat{\beta}^{(-)}=\frac{-k_2\beta'-k_2'\al}{G_1^{(-)}G_2^{(-)}}.
\ee
In these expressions, $X$ and $Y$ can be either $S$ or $O$, and we make use of the definitions (\ref{eq:GDefs}), (\ref{eq:PlusDefs}), and (\ref{eq:MinusDefs}).

What remains is the fusion of two non-generic non-factorized $\mcO^{OO}$ interfaces.  These can be worked out on a case by case basis.  In the next section we will compute the fusion algebra of the topological interfaces for a theory at a fixed radius.

\subsection{Identification of interesting non-factorized interfaces}

The only $S^1|S^1/\Z_2$ interfaces of real interest for us are the ``orbifold interfaces''.  These are $\mcO^{(+)\,SO}_{1,1}(0;0)$ and $\mcO^{(+)\,OS}_{1,1}(0;0)$ for the case $R_1=R_2$.  These interfaces are topological, and implement the isomorphism between invariant states of the circle theory and untwisted sector states of the orbifold theory.  Indeed, we have
\begin{multline}
\mcO^{(+)\,OS}_{1,1}(0;0)\prod_{n=1}^\infty\lp a_n^{2\,\dagger}\rp^{m_n}\lp\wta_n^{2\,\dagger}\rp^{\bar{m}_n}|N,M\rangle_2=\frac{1}{\sqrt{2}}\prod_{n=1}^\infty\lp a_n^{2\,\dagger}\rp^{m_n}\lp\wta_n^{2\,\dagger}\rp^{\bar{m}_n}|N,M\rangle_2\\
+\frac{1}{\sqrt{2}}\prod_{n=1}^\infty\lp -a_n^{2\,\dagger}\rp^{m_n}\lp -\wta_n^{2\,\dagger}\rp^{\bar{m}_n}|-N,-M\rangle_2,
\end{multline}
so this interface projects onto invariant states and sends them to the untwisted sector of the orbifold.  Also,
\be
\mcO^{(+)\,SO}_{1,1}(0;0)\circ\mcO^{(+)\,OS}_{1,1}(0;0)=\mcO^{(+)}_{1,1}(0;0)+\mcO^{(+)}_{1,-1}(0;0),
\ee
which is the sum of the identity interface and the reflection interface, while
\be
\mcO^{(+)\,OS}_{1,1}(0;0)\circ\mcO^{(+)\,SO}_{1,1}(0;0)=\mcO^{(+)\,OO}_{1,1}(0;0;+)+\mcO^{(+)\,OO}_{1,1}(0;0;-),
\ee
which projects onto untwisted sector states.

Let's take a moment here to point out another case where we are missing interfaces that we expect, just like the case of the missing symmetry interfaces in the $S^1$ theory at the self-dual radius.  It is known that the $S^1/\Z_2$ orbifold at the self-dual radius $R_2=1/\sqrt{2}$ is isomorphic, i.e.\ physically equivalent, to the $S^1$ theory at twice the self-dual radius, $R_1=\sqrt{2}$.  There should then be an invertible topological interface which implements this isomorphism.  However we can easily find all the topological interfaces in this case,
\be
\mcO^{(+)\,SO}_{1,2}(\al;\beta),\qquad\mcO^{(-)\,SO}_{2,1}(\al;\beta),
\ee
or going the other way,
\be
\mcO^{(+)\,OS}_{2,1}(\al;\beta),\qquad\mcO^{(-)\,OS}_{2,1}(\al;\beta).
\ee
However, it is easy to check that these interfaces are not invertible.  In fact, they don't even see the twisted sector states on the orbifold side at all, so they certainly can't be the isomorphism interfaces we expect (which should map twisted sector states into non-invariant states in the $S^1$ theory).

It should not come as a surprise that we are missing this interface, given that we were missing some of the enhanced global symmetries at the self-dual radius.  The easiest way to prove the isomorphism between $S^1/\Z_2$ at $R=1/\sqrt{2}$ and $S^1$ at $R=\sqrt{2}$ is to notice that they are both given by $\Z_2$ orbifolds of the self-dual $S^1$ theory, and that the two $\Z^2$ subgroups in question are conjugate inside the full $\SU(2)^2/\Z_2$ global symmetry group.  Since we are missing the interfaces that implement the conjugation, we shouldn't be surprised that we are missing the interface which implements the isomorphism.

Moving on to $(S^1/\Z_2)|(S^1/\Z_2)$ interfaces, we will restrict ourselves to computing the fusion algebra of the topological interfaces for a theory at a fixed radius.  For generic radius, these will exclusively be the $\mcI^{(+)\,OO}_{1,1}$ interfaces.  We find
\bea
\mcI^{(+)\,OO}_{1,1}(\al;\beta)\circ\mcI^{(+)\,OO}_{1,1}(\al';\beta') &=& \mcI^{(+)\,OO}_{1,1}(\al+\al';\beta+\beta')+\mcI^{(+)\,OO}_{1,1}(\al-\al';\beta-\beta'),\non\\
\\
\mcI^{(+)\,OO}_{1,1}(\al;\beta)\circ\mcI^{(+)\,OO}_{1,1}(\al_0';\beta_0';\e') &=& \mcI^{(+)\,OO}_{1,1}(\al+\al_0';\beta+\beta_0'),\\
\mcI^{(+)\,OO}_{1,1}(\al_0;\beta_0;\e)\circ\mcI^{(+)\,OO}_{1,1}(\al';\beta') &=& \mcI^{(+)\,OO}_{1,1}(\al_0+\al';\beta_0+\beta'),\\
\mcI^{(+)\,OO}_{1,1}(\al_0;\beta_0;\e)\circ\mcI^{(+)\,OO}_{1,1}(\al_0';\beta_0';\e') &=& \mcI^{(+)\,OO}_{1,1}(\al_0+\al_0';\beta_0+\beta_0';(-1)^{\frac{\beta_0\al_0'}{\pi^2}}\e\e').
\eea
We see that the generic interfaces are topological but not invertible.  There are eight invertible interfaces, $\mcI^{(+)\,OO}_{1,1}(\al_0;\beta_0;\e)$, that generate a global symmetry group $D_4$, the dihedral group of order eight.  To verify this, we note that $\mcI^{(+)\,OO}_{1,1}(0;0;+)=e$ acts as the identity\footnote{From the form of $|v^{(+)}_{1,1}(0;0)\rangle^t$, we can see that this interface indeed represents an insertion of a complete set of normalized states of the orbifold theory.}, and if we define $a=\mcI^{(+)\,OO}_{1,1}(0;\pi;+)$, $b=\mcI^{(+)\,OO}_{1,1}(\pi;\pi;+)$, then we can verify that $a$ has order two, $b$ has order four, and that $aba=b^3$.

At the self-dual radius, we have additional topological interfaces $\mcI^{(-)}_{1,1}$.  Focusing only on the non-generic ones, we compute
\bea
\mcI^{(+)\,OO}_{1,1}(\al_0;\beta_0;\e)\circ\mcI^{(-)\,OO}_{1,1}(\al_0';\beta_0';\e') &=& \mcI^{(-)\,OO}_{1,1}(\al_0'+\beta_0;\beta_0'+\al_0;(-1)^{\frac{\beta_0\beta_0'}{\pi^2}}\e\e'),\\
\mcI^{(-)\,OO}_{1,1}(\al_0;\beta_0;\e)\circ\mcI^{(+)\,OO}_{1,1}(\al_0';\beta_0';\e') &=& \mcI^{(-)\,OO}_{1,1}(\al_0+\al_0';\beta_0+\beta_0';(-1)^{\frac{\beta_0'(\al_0+\al_0')}{\pi^2}}\e\e'),\\
\mcI^{(-)\,OO}_{1,1}(\al_0;\beta_0;\e)\circ\mcI^{(-)\,OO}_{1,1}(\al_0';\beta_0';\e') &=& \mcI^{(+)\,OO}_{1,1}(\al_0'+\beta_0;\beta_0'+\al_0;(-1)^{\frac{\al_0'(\al_0+\beta_0')}{\pi^2}}\e\e').
\eea
These are all invertible and extend the realized global symmetry group to $D_8$, the dihedral group of order sixteen.  It should not be a surprise at this point that this is still not the full symmetry group for the $S^1/\Z_2$ theory at the self-dual radius, which is expected to be enhanced to $(U(1)\rtimes\Z_2)^2$.

Lastly, let us note that the invertible non-topological interface $\mcO^{(+)\,OO}_{1,1}(0;0;+)$, for the case where $R_2\ne R_1$, has the correct behavior to be the radius-changing interface, i.e.\ the interface which implements the marginal deformation from the $S^1/\Z_2$ theory at radius $R_2$ to the theory at radius $R_1$.  Specifically, it reduces to the identity interface when $R_2=R_1$, and it composes correctly under fusion.

\section{Remarks and future directions}
\label{sec:Conclusions}

We have constructed a large class of conformal interfaces between $c=1$ theories, and there have implicitly been many checks along the way - in particular the fact that our fusion algebra involved only products that were nonnegative integer combinations of physical interfaces is a strong indication that our constructions were consistent.

However, we have also noted that there are some interfaces still missing, such as the enhanced symmetries at the self-dual radius or the isomorphism between the $S^1/\Z_2$ theory at the self-dual radius and the $S^1$ theory at twice the self-dual radius.  Related to these, there should also be invertible interfaces between any $S^1/\Z_2$ theory and any $S^1$ theory corresponding to the marginal deformation which passes through that point.  It would be very nice to construct these missing pieces and hopefully get a complete classification of possible interfaces between $c=1$ theories, perhaps also including the isolated CFTs of the Ginsparg archipelago~\cite{Ginsparg:1988ui,Ginsparg:1987eb}.  For supersymmetric $c=3/2$ models, it would be interesting to extend the results of~\cite{Bachas:2012bj} to the other branches of theories found in~\cite{Dixon:1988ac}.

Another direction of investigation would be to understand the Casimir forces between these interfaces, or between interfaces and boundary conditions, and look for possible attractor mechanisms~\cite{Brunner:2010xm, bachas02}.

Finally, it would be especially interesting to extend the algebra of conformal interfaces to $c\le 1$, i.e.\ to include also the cases where one of the CFTs can be a minimal model, and in particular to understand the RG interfaces governing flow from the $c=1$ theories to the $c<1$ theories, and how those interfaces fuse with the ones studied in this paper.  This might in turn offer a new perspective on the space of two-dimensional conformal field theories and the connections between them, along the lines of~\cite{Gukov:2015qea}.

\section*{Acknowledgements}

D.~R.~would like to thank Oleg Lunin for useful discussion and comments.  This work was performed in part at the Aspen Center for Physics, which is supported by the National Science Foundation under grant PHY-1607611.
This work was also supported by NSF under grants PHY-1521099 and PHY-1620742 and the Mitchell Institute for Fundamental Physics and Astronomy.

\appendix
\section{Theta functions}
\label{app:ThetaFunctions}

We use the following conventions for theta functions, where $q=e^{2\pi i\tau}$, $y=e^{2\pi iz}$,
\bea
\vartheta_1(\tau,z) &=& -i\sum_{n\in\Z}\lp -1\rp^ny^{n+\hlf}q^{\hlf\lp n+\hlf\rp^2}=2q^{\frac{1}{8}}\sin(\pi z)\prod_{m=1}^\infty\lp 1-q^m\rp\lp 1-yq^m\rp\lp 1-y^{-1}q^m\rp,\non\\
&& \\
\vartheta_2(\tau,z) &=& \sum_{n\in\Z}y^{n+\hlf}q^{\hlf\lp n+\hlf\rp^2}=2q^{\frac{1}{8}}\cos(\pi z)\prod_{m=1}^\infty\lp 1-q^m\rp\lp 1+yq^m\rp\lp 1+y^{-1}q^m\rp,\\
\vartheta_3(\tau,z) &=& \sum_{n\in\Z}y^nq^{\hlf n^2}=\prod_{m=1}^\infty\lp 1-q^m\rp\lp 1+yq^{m-\hlf}\rp\lp 1+y^{-1}q^{m-\hlf}\rp,\\
\vartheta_4(\tau,z) &=& \sum_{n\in\Z}\lp -1\rp^ny^nq^{\hlf n^2}=\prod_{m=1}^\infty\lp 1-q^m\rp\lp 1-yq^{m-\hlf}\rp\lp 1-y^{-1}q^{m-\hlf}\rp.
\eea
We will often use the theta function evaluated at $z=0$, in which case we will usually omit that argument, $\vartheta_i(\tau)=\vartheta(\tau,0)$.  We will also be sloppy in passing between using $\tau$ as the function argument versus using $q$, relying on context to differentiate between the two possibilities.

The Dedekind eta function is defined as
\be
\eta(\tau)=q^{\frac{1}{24}}\prod_{m=1}^\infty\lp 1-q^m\rp.
\ee

Under the transformation $\tau\rr -1/\tau$, these functions transform as
\bea
\eta(-\frac{1}{\tau}) &=& \sqrt{-i\tau}\eta(\tau),\\
\vartheta_1(-\frac{1}{\tau},\frac{z}{\tau}) &=& -ie^{\pi i\frac{z^2}{\tau}}\sqrt{-i\tau}\vartheta_1(\tau,z),\\
\vartheta_2(-\frac{1}{\tau},\frac{z}{\tau}) &=& e^{\pi i\frac{z^2}{\tau}}\sqrt{-i\tau}\vartheta_4(\tau,z),\\
\vartheta_3(-\frac{1}{\tau},\frac{z}{\tau}) &=& e^{\pi i\frac{z^2}{\tau}}\sqrt{-i\tau}\vartheta_3(\tau,z),\\
\vartheta_4(-\frac{1}{\tau},\frac{z}{\tau}) &=& e^{\pi i\frac{z^2}{\tau}}\sqrt{-i\tau}\vartheta_2(\tau,z).
\eea

\section{Annulus computations}
\label{app:Annulus}

\subsection{$S^1$ computations}

The annulus computations generally split into an oscillator part and a zero-mode part.  For the oscillators in the circle theory, the most general possibility involves a product over $n\in\Z$ of
\be
\mathcal{A}_n^{(\e\e')}=\langle 0|e^{\e a_n\wta_n}q^{n\lp a_n^\dagger a_n+\wta_n^\dagger\wta_n\rp}e^{\e'a_n^\dagger\wta_n^\dagger}|0\rangle,
\ee
where $\e$ and $\e'$ are $\pm$.  To compute this, we first note that $a_n^\dagger a_n$ and $\wta_n^\dagger\wta_n$ simply count the number of $a_n^\dagger$ or $\wta_n^\dagger$ raising operators, i.e.\
\be
a_n^\dagger a_n\lp a_n^\dagger\rp^k|0\rangle=k\lp a_n^\dagger\rp^k|0\rangle,\qquad\wta_n^\dagger\wta_n\lp\wta_n^\dagger\rp^k=k\lp\wta_n^\dagger\rp^k|0\rangle,
\ee
and thus
\be
\label{eq:Absorbq}
q^{n\lp a_n^\dagger a_n+\wta_n^\dagger\wta_n\rp}e^{\e'a_n^\dagger\wta_n^\dagger}|0\rangle=q^{n\lp a_n^\dagger a_n+\wta_n^\dagger\wta_n\rp}\sum_{k=0}^\infty\frac{(\e')^k}{k!}\lp a_n^\dagger\wta_n^\dagger\rp^k|0\rangle=\e^{\e'q^{2n}a_n^\dagger\wta_n^\dagger}|0\rangle.
\ee
Next we observe that
\be
\langle 0|\lp a_n\wta_n\rp^k\lp a_n^\dagger\wta_n^\dagger\rp^\ell|0\rangle=\d_{k\ell}c_k,
\ee
for some constants $c_k$ (if $k\ne\ell$ then we can move the extra operators through to annihilate the corresponding vacuum).  Proceeding by induction,
\bea
c_k &=& \langle 0|\lp a_n\wta_n\rp^{k-1}a_n\ls\wta_n,\lp a_n^\dagger\wta_n^\dagger\rp^k\rs|0\rangle\non\\
&=& k\langle 0|\lp a_n\wta_n\rp^{k-1}\ls a_n\lp a_n^\dagger\rp^k\rs\lp\wta_n^\dagger\rp^{k-1}|0\rangle\non\\
&=& k^2\langle 0|\lp a_n\wta_n\rp^{k-1}\lp a_n^\dagger\wta_n^\dagger\rp^{k-1}|0\rangle=k^2c_{k-1}.
\eea
Combining this with the fact that $c_0=1$, we find that $c_k=(k!)^2$, and finally
\be
\mathcal{A}_n^{(\e\e')}=\sum_{k,\ell=0}^\infty\frac{\e^k(\e'q^{2n})^\ell}{k!\ell!}\langle 0|\lp a_n\wta_n\rp^k\lp a_n^\dagger\wta_n^\dagger\rp^\ell|0\rangle=\sum_{k=0}^\infty\lp\e\e'q^{2n}\rp^k=\frac{1}{1-\e\e'q^{2n}}.
\ee
In particular
\be
\prod_{n=1}^\infty\mathcal{A}_n^{(\e\e')}=\left\{\begin{matrix}\frac{q^{\frac{1}{12}}}{\eta(q^2)}, & \mathrm{if\ }\e=\e',\\ q^{\frac{1}{12}}\sqrt{\frac{2\eta(q^2)}{\vartheta_2(q^2)}}, & \mathrm{if\ }\e=-\e'.\end{matrix}\right.
\ee

Adding the zero-mode pieces, we get
\bea
\label{eq:AppANN}
\langle\langle N(\al)|q^H|N(\al')\rangle\rangle &=& R\lp\prod_{n=1}^\infty\mathcal{A}_n^{(--)}\rp\sum_{M,M'\in\Z}e^{iM\al-iM'\al'}\langle 0,M|q^{(M')^2R^2-\frac{1}{12}}|0,M'\rangle\non\\
&=& \frac{R}{\eta(q^2)}\sum_{M\in\Z}e^{iM(\al-\al')}q^{M^2R^2},\\
\label{eq:AppAND}
\langle\langle N(\al)|q^H|D(\beta')\rangle\rangle &=& \frac{1}{\sqrt{2}}\lp\prod_{n=1}^\infty\mathcal{A}_n^{(-+)}\rp\sum_{M,N'\in\Z}e^{iM\al-iN'\beta'}\langle 0,M|q^{\frac{(N')^2}{4R^2}-\frac{1}{12}}|N',0\rangle\non\\
&=& \sqrt{\frac{\eta(q^2)}{\vartheta_2(q^2)}},\\
\label{eq:AppADD}
\langle\langle D(\beta)|q^H|D(\beta')\rangle\rangle &=& \frac{1}{2R}\lp\prod_{n=1}^\infty\mathcal{A}_n^{(++)}\rp\sum_{N,N'\in\Z}e^{iN\beta-iN'\beta'}\langle N,0|q^{\frac{(N')^2}{4R^2}-\frac{1}{12}}|N',0\rangle\non\\
&=& \frac{1}{2R\eta(q^2)}\sum_{N\in\Z}e^{iN(\beta-\beta')}q^{\frac{N^2}{4R^2}},
\eea
which gives (\ref{eq:ANN})-(\ref{eq:ADD}).

Finally we need to perform the modular transformation from $q=e^{-T}$ to $\wtq=e^{-\pi^2/T}$.  From Appendix \ref{app:ThetaFunctions} we have (hopefully the reader is not too confused as we go back and forth between $\eta(q)$ and $\eta(\tau)$)
\be
\eta(q^2)=\eta(\frac{iT}{\pi})=\sqrt{\frac{\pi}{T}}\eta(\frac{i\pi}{T})=\sqrt{\frac{\pi}{T}}\eta(\wtq^2),\qquad\vartheta_2(q^2)=\sqrt{\frac{\pi}{T}}\vartheta_4(\wtq^2).
\ee
This is enough to get the mixed annulus,
\be
\langle\langle N(\al)|q^H|D(\beta')\rangle\rangle=\sqrt{\frac{\eta(\wtq^2)}{\vartheta_4(\wtq^2)}},
\ee
as in (\ref{eq:ATND}).  For the other two, we also need to perform a Poisson resummation on the zero modes.  The general result for of the resummation that we will need is
\be
\sum_{n\in\Z}e^{-\pi an^2+2\pi ibn}=\frac{1}{\sqrt{a}}\sum_{m\in\Z}e^{-\frac{\pi(m-b)^2}{a}}.
\ee
Applying this to the zero mode sums, we get
\be
\sum_{M\in\Z}e^{iM(\al-\al')}q^{M^2R^2}=\frac{1}{R}\sqrt{\frac{\pi}{T}}\sum_{M\in\Z}\wtq^{\frac{1}{R^2}(M-\frac{\al-\al'}{2\pi})^2},
\ee
\be
\sum_{N\in\Z}e^{iN(\beta-\beta')}q^{\frac{N^2}{4R^2}}=2R\sqrt{\frac{\pi}{T}}\sum_{N\in\Z}\wtq^{4R^2(N-\frac{\beta-\beta'}{2\pi})^2}.
\ee
Substituting these in (\ref{eq:AppANN}) and (\ref{eq:AppADD}) gives us (\ref{eq:ATNN}) and (\ref{eq:ATDD}).

\subsection{$S^1/\Z_2$ computations}
\label{app:OrbifoldAnnulus}

In the orbifold, the twisted sector computations are nearly identical, with
\be
\mathcal{B}_r^{(\e\e')}=\langle 0|e^{\e a_r^t\wta_r^t}q^{r\lp a_r^{t\,\dagger}a_r^t+\wta_r^{t\,\dagger}\wta_r^t\rp}e^{\e'a_r^{t\,\dagger}\wta_r^{t\,\dagger}}|0\rangle=\frac{1}{1-\e\e'q^{2r}}.
\ee
Then
\bea
\vphantom{\langle}^t\langle\langle N_O(\al)|q^{H^t}|N_O(\al')\rangle\rangle^t &=& \d_{\al\al'}q^{\frac{1}{24}}\lp\prod_{r\in\Z+\hlf,r>0}\mathcal{B}_r^{(--)}\rp=\d_{\al\al'}\sqrt{\frac{\eta(q^2)}{\vartheta_4(q^2)}},\non\\
\vphantom{\langle}^t\langle\langle N_O(\al)|q^{H^t}|D_O(\beta')\rangle\rangle^t &=& \lp -1\rp^{\frac{\al\beta'}{\pi^2}}\frac{q^{\frac{1}{24}}}{\sqrt{2}}\lp\prod_{r\in\Z+\hlf,r>0}\mathcal{B}_r^{(-+)}\rp=\lp -1\rp^{\frac{\al\beta'}{\pi^2}}\sqrt{\frac{\eta(q^2)}{2\vartheta_3(q^2)}},\\
\vphantom{\langle}^t\langle\langle D_O(\beta)|q^{H^t}|D_O(\beta')\rangle\rangle^t &=& \d_{\beta\beta'}q^{\frac{1}{24}}\lp\prod_{r\in\Z+\hlf,r>0}\mathcal{B}_r^{(++)}\rp=\d_{\beta\beta'}\sqrt{\frac{\eta(q^2)}{\vartheta_4(q^2)}}.
\eea

Then taking the combinations (\ref{eq:OrbB1})-(\ref{eq:OrbB4}) we can compute all of the modular transformed annulus amplitudes,
\bea
\langle\langle N_O(\al)|q^H|N_O(\al'\rangle\rangle &=& \frac{1}{\eta(\wtq^2)}\lp\sum_{M\in\Z}\wtq^{\frac{1}{R^2}(M+\frac{\al'-\al}{2\pi})^2}+\sum_{M\in\Z}\wtq^{\frac{1}{R^2}(M+\frac{\al'+\al}{2\pi})^2}\rp,\\
\langle\langle N_O(\al)|q^H|N_O(\al_0',\e')\rangle\rangle &=& \frac{1}{\eta(\wtq^2)}\sum_{M\in\Z}\wtq^{\frac{1}{R^2}(M+\frac{\al-\al_0'}{2\pi})^2},\\
\langle\langle N_O(\al)|q^H|D_O(\beta')\rangle\rangle &=& 2\sqrt{\frac{\eta(\wtq^2)}{\vartheta_4(\wtq^2)}},\\
\langle\langle N_O(\al)|q^H|D_O(\beta_0',\e')\rangle\rangle &=& \sqrt{\frac{\eta(\wtq^2)}{\vartheta_4(\wtq^2)}},\\
\langle\langle N_O(\al_0,\e)|q^H|N_O(\al_0',\e')\rangle\rangle &=& \frac{1}{2\eta(\wtq^2)}\sum_{M\in\Z}\wtq^{\frac{1}{R^2}(M+\frac{\al_0'-\al_0}{2\pi})^2}+\e\e'\d_{\al_0\al_0'}\sqrt{\frac{\eta(\wtq^2)}{2\vartheta_2(\wtq^2)}},\\
\langle\langle N_O(\al_0,\e)|q^H|D_O(\beta')\rangle\rangle &=& \sqrt{\frac{\eta(\wtq^2)}{\vartheta_4(\wtq^2)}},\\
\langle\langle N_O(\al_0,\e)|q^H|D_O(\beta_0',\e')\rangle\rangle &=& \hlf\sqrt{\frac{\eta(\wtq^2)}{\vartheta_4(\wtq^2)}}+\hlf\e\e'\lp -1\rp^{\frac{\al_0\beta_0'}{\pi^2}}\sqrt{\frac{\eta(\wtq^2)}{\vartheta_3(\wtq^2)}},\\
\langle\langle D_O(\beta)|q^H|D_O(\beta')\rangle\rangle &=& \frac{1}{\eta(\wtq^2)}\lp\sum_{N\in\Z}\wtq^{4R^2(N+\frac{\beta'-\beta}{2\pi})^2}+\sum_{N\in\Z}\wtq^{4R^2(N+\frac{\beta'+\beta}{2\pi})^2}\rp,\\
\langle\langle D_O(\beta)|q^H|D_O(\beta_0',\e')\rangle\rangle &=& \frac{1}{\eta(\wtq^2)}\sum_{N\in\Z}\wtq^{4R^2(N+\frac{\beta-\beta_0'}{2\pi})^2},\\
\langle\langle D_O(\beta_0,\e)|q^H|D_O(\beta_0',\e')\rangle\rangle &=& \frac{1}{2\eta(\wtq^2)}\sum_{N\in\Z}\wtq^{4R^2(N+\frac{\beta_0'-\beta_0}{2\pi})^2}+\e\e'\d_{\beta_0\beta_0'}\sqrt{\frac{\eta(\wtq^2)}{2\vartheta_2(\wtq^2)}},
\eea
and it is easily verified that these are thus all mutually Cardy consistent.

\subsection{$(S^1)^2$ computations}
\label{app:T2Ann}

For the oscillator part of an annulus computation in $(S^1)^2$, we need to compute
\be
\mathcal{A}_n^{(\e\e')}(\theta,\theta')=\langle 0|e^{a_n^T\cdot S\cdot\wta_n}q^{n\lp a_n^{\dagger\,T}a_n+\wta_n^{\dagger\,T}\wta_n\rp}e^{a_n^{\dagger\,T}\cdot S'\cdot\wta_n^\dagger}|0\rangle,
\ee
where we have assembled oscillators into two-vectors, $a_n^T=(a_n^1,a_n^2)$, etc., and where
\be
S=\lp\begin{matrix} -\e\cos(2\theta) & -\sin(2\theta) \\ -\e\sin(2\theta) & \cos(2\theta)\end{matrix}\rp,\qquad S'=\lp\begin{matrix} -\e'\cos(2\theta') & -\sin(2\theta') \\ -\e'\sin(2\theta') & \cos(2\theta')\end{matrix}\rp,
\ee
where $\e$ and $\e'$ are $\pm$, and without loss of generality $0\le\theta,\theta'<\pi$.

First we note that, by the same arguments as in the $S^1$ case,
\be
q^{n\lp a_n^{\dagger\,T}a_n+\wta_n^{\dagger\,T}\wta_n\rp}e^{a_n^{\dagger\,T}\cdot S'\cdot\wta_n^\dagger}|0\rangle=e^{q^{2n}a_n^{\dagger\,T}\cdot S'\cdot\wta_n^\dagger}|0\rangle.
\ee

Next, let us define
\be
A=a_n^T\cdot S\cdot\wta_n,\qquad B=a_n^{\dagger\,T}\cdot S'\cdot\wta_n^\dagger,
\ee
and consider
\be
\langle 0|A^kB^\ell|0\rangle=\d_{k\ell}c_k.
\ee
After some algebra we get
\bea
c_k &=& \langle 0|A^{k-1}\ls A,B^k\rs|0\rangle\non\\
&=& k\langle 0|A^{k-1}\left\{\lp -\e\cos(2\theta)a_n^1-\e\sin(2\theta)a_n^2\rp\lp -\e'\cos(2\theta')a_n^{1\,\dagger}-\e'\sin(2\theta')a_n^{2\,\dagger}\rp\right.\non\\
&& \quad\left. +\lp -\sin(2\theta)a_n^1+\cos(2\theta)a_n^2\rp\lp -\sin(2\theta')a_n^{1\,\dagger}+\cos(2\theta')a_n^{2\,\dagger}\rp\right\}B^{k-1}|0\rangle\non\\
&=& \langle 0|A^{k-1}\left\{-k(k-1)\e\e'A^\dagger+k^2\lp 1+\e\e'\rp\cos(2(\theta'-\theta))B\right\}B^{k-2}|0\rangle.
\eea
Using the fact that
\be
\ls A,A^\dagger\rs=a_n^1a_n^{1\,\dagger}+a_n^2a_n^{2\,\dagger}+\wta_n^1\wta_n^{1\,\dagger}+\wta_n^2\wta_n^{2\,\dagger}=2+N_{tot},
\ee
where $N_{tot}$ is the number operator, we can show that
\be
\langle 0|A^{k-1}A^\dagger=k(k-1)\langle 0|A^{k-2}.
\ee
Thus, we get the recursion relation
\be
c_k=k^2\lp 1+\e\e'\rp\cos(2(\theta'-\theta))c_{k-1}-\e\e'k^2(k-1)^2c_{k-2}.
\ee
With initial values $c_0=1$ and $c_1=(1+\e\e')\cos(2(\theta'-\theta))$, we can solve this recursion.  We find
\be
c_k=\left\{\begin{matrix}\lp k!\rp^2\frac{\sin(2(k+1)(\theta'-\theta))}{\sin(2(\theta'-\theta))}, & \e'=\e,\\ \lp k!\rp^2\frac{1+\lp -1\rp^k}{2}, & \e'=-\e.\end{matrix}\right.
\ee
With these results, we get
\be
\mathcal{A}_n^{(\e\e')}(\theta,\theta')=\sum_{k=0}^\infty\frac{q^{2nk}}{(k!)^2}c_k=\left\{\begin{matrix}\frac{1}{\lp 1-e^{2i(\theta'-\theta)}q^{2n}\rp\lp 1-e^{-2i(\theta'-\theta)}q^{2n}\rp}, & \e'=\e,\\ \frac{1}{1-q^{4n}}, & \e'=-\e.\end{matrix}\right.
\ee

Now for the full amplitude we combine these results with the sum over the zero-modes.  We'll start with the case in which both boundary states are ND, so $\e=\e'=+1$.  In this case the annulus diagram is (dropping the overall normalization factors $g$ and $g'$)
\begin{multline}
\langle\langle B^{(+)}_{k_1,k_2}(\al;\beta)|q^H|B^{(+)}_{k_1',k_2'}(\al';\beta')\rangle\rangle\\
=q^{-\frac{1}{6}}\lp\prod_{n=1}^\infty\mathcal{A}_n^{(++)}(\vartheta,\vartheta')\rp\sum_{N,M,N',M'\in\Z}\d_{k_1N,k_1'N'}\d_{k_1M,k_1'M'}\d_{k_2N,k_2'N'}\d_{k_2M,k_2'M'}\\
\times e^{iM\al-iM'\al'+iN\beta-iN'\beta'}q^{\lp (k_1'R_1)^2+(k_2'R_2)^2\rp\lp\frac{(N')^2}{4R_1^2R_2^2}+(M')^2\rp}.
\end{multline}

If $(k_1',k_2')=(k_1,k_2)$ (so also $\vartheta'=\vartheta$), then the Kronecker deltas simply enforce $N'=N$ and $M'=M$ and the above result becomes
\be
\frac{1}{\eta(q^2)^2}\sum_{M,N\in\Z}e^{iM(\al-\al')+iN(\beta-\beta')}q^{\lp k_1^2R_1^2+k_2^2R_2^2\rp\lp\frac{N^2}{4R_1^2R_2^2}+M^2\rp}.
\ee
Performing a modular transformation, including Poisson resummation, leads to
\be
\frac{2R_1R_2}{k_1^2R_1^2+k_2^2R_2^2}\frac{1}{\eta(\wtq^2)^2}\sum_{M,N\in\Z}\wtq^{\frac{4R_1^2R_2^2\lp N-\frac{\beta-\beta'}{2\pi}\rp^2+\lp M-\frac{\al-\al'}{2\pi}\rp^2}{k_1^2R_1^2+k_2^2R_2^2}},
\ee
as in (\ref{eq:ANDND1}).  If $(k_1',k_2')=(-k_1,-k_2)$, then we can use the equivalence (\ref{eq:Redundancy}).

Otherwise (recalling that $(k_1,k_2)$ and $(k_1',k_2')$ are relatively prime pairs), it must be that the ratios $k_1:k_1'$ and $k_2:k_2'$ are not equal, in which case the Kronecker deltas enforce $N=N'=M=M'=0$.  In this case the annulus gives
\be
\frac{2\sin(\vartheta'-\vartheta)\eta(q^2)}{\vartheta_1(q^2,\frac{\vartheta'-\vartheta}{\pi})}=\frac{2i\sin(\vartheta'-\vartheta)\wtq^{\,-\lp\frac{\vartheta'-\vartheta}{\pi}\rp^2}\eta(\wtq^2)}{\vartheta_1(\wtq^2,i\frac{\vartheta'-\vartheta}{T})}.
\ee
Finally, using
\begin{multline}
\sin(\vartheta'-\vartheta)=\cos\vartheta\sin\vartheta'-\sin\vartheta\cos\vartheta'\\
=\lp\frac{k_1R_1}{\sqrt{k_1^2R_1^2+k_2^2R_2^2}}\rp\lp\frac{k_2'R_2}{\sqrt{(k_1'R_1)^2+(k_2'R_2)^2}}\rp-\lp\frac{k_2R_2}{\sqrt{k_1^2R_1^2+k_2^2R_2^2}}\rp\lp\frac{k_1'R_1}{\sqrt{(k_1'R_1)^2+(k_2'R_2)^2}}\rp,
\end{multline}
we recover (\ref{eq:ANDND2}).

The annulus diagram with two $B^{(-)}$ states is very similar, and can be obtained from the above by T-duality, so the last annulus we need to compute is
\begin{multline}
\langle\langle B^{(+)}_{k_1,k_2}(\al;\beta)|q^H|B^{(-)}_{k_1',k_2'}(\al';\beta')\rangle\rangle\\
=gg'q^{-\frac{1}{6}}\lp\prod_{n=1}^\infty\mathcal{A}_n^{(+-)}(\theta,\theta')\rp\sum_{N,M,N',M'\in\Z}\d_{k_2N,-k_1'M'}\d_{k_1M,-k_2'N'}\d_{k_1N,k_1'N'}\d_{k_2M,k_2'M'}\\
\times e^{iM\al-iM'\al'+iN\beta-iN'\beta'}q^{\lp(k_1')^2+4(k_2'R_1R_2)^2\rp\lp\frac{(N')^2}{4R_2^2}+\frac{(M')^2}{4R_1^2}\rp}.
\end{multline}
The Kronecker deltas enforce
\be
N=k_1'L,\quad M=-k_2'L,\quad N'=k_1L,\quad M'=-k_2L,\qquad L\in\Z.
\ee
Thus the annulus becomes
\be
\frac{\sqrt{2}gg'}{\sqrt{\eta(q^2)\vartheta_2(q^2)}}
\sum_{L\in\Z}e^{i\lp k_2\al'-k_2'\al-k_1\beta'+k_1'\beta\rp L}q^{\frac{\lp(k_1')^2+4(k_2'R_1R_2)^2\rp\lp k_1^2R_1^2+k_2^2R_2^2\rp}{4R_1^2R_2^2}L^2}.
\ee
After modular transformation, we get
\be
\frac{1}{\sqrt{\eta(\wtq^2)\vartheta_4(\wtq^2)}}\sum_{L\in\Z}\wtq^{\frac{4R_1^2R_2^2\lp L-\frac{1}{2\pi}\lp k_2\al'-k_2'\al-k_1\beta'+k_1'\beta\rp\rp^2}{\lp k_1^2R_1^2+k_2^2R_2^2\rp\lp(k_1')^2+4(k_2'R_1R_2)^2\rp}},
\ee
exactly as in (\ref{eq:ANDDD}).

\subsection{$(S^1/\Z_2)^2$ computations}

In the orbifold we don't need much more than we have already computed in the $(S^1)^2$ and $S^1/\Z_2$ sections.  The only new part which needs some discussion is the twisted-sector oscillator part.  If we define
\be
\mathcal{B}_r^{(\e\e')}(\theta,\theta')=\langle 0|e^{a_r^{t\,T}\cdot S^{(\e)}\cdot\wta_r^t}q^{r\lp a_r^{t\,\dagger\,T}a_r^t+\wta_r^{t\,\dagger\,T}\wta_r^t\rp}e^{a_r^{t\,\dagger\,T}\cdot S^{(\e')}\cdot\wta_r^{t\,\dagger}}|0\rangle.
\ee
Then the same manipulations as in the previous section give us
\be
\mathcal{B}_r^{(\e\e')}(\theta,\theta')=\left\{\begin{matrix}\frac{1}{\lp 1-e^{2i(\theta'-\theta)}q^{2r}\rp\lp 1-e^{-2i(\theta'-\theta)}q^{2r}\rp}, & \e'=\e,\\ \frac{1}{1-q^{4r}}, & \e'=-\e.\end{matrix}\right.
\ee
And then, if $\e'=\e$, we have
\be
\prod_{r\in\Z+\hlf,r>0}\mathcal{B}_r^{(\pm\pm)}(\theta',\theta)=q^{-\frac{1}{12}}\frac{\eta(q^2)}{\vartheta_4(q^2,\frac{\theta'-\theta}{\pi})},
\ee
and if $\e'=-\e$,
\be
\prod_{r\in\Z+\hlf,r>0}\mathcal{B}_r^{(\pm\mp)}(\theta',\theta)=q^{-\frac{1}{12}}\frac{\eta(q^2)}{\sqrt{\vartheta_3(q^2)\vartheta_4(q^2)}}.
\ee
Multiplying by the zero-mode contribution $q^{\frac{1}{12}}$ and performing a modular transformation, we get
\be
\frac{\wtq^{\,-\lp\frac{\theta'-\theta}{\pi}\rp^2}\eta(\wtq^2)}{\vartheta_2(\wtq^2,i\frac{\theta'-\theta}{T})},\qquad\mathrm{and}\qquad\frac{\eta(\wtq^2)}{\sqrt{\vartheta_2(\wtq^2)\vartheta_3(\wtq^2)}},
\ee
respectively.

\section{Fusion computations}
\label{app:Fusion}

\subsection{$S^1$ interface fusion}
\label{subapp:S1Fusion}

We would like to compute the fusion of two general interfaces between $S^1$ theories, one taking operators in a theory at radius $R_2$ to operators in a theory at radius $R_1$, and one going from $R_3$ to $R_2$, finally obtaining an interface going from the $R_3$ theory to the $R_1$ theory.  The computation is formulated in section \ref{subsec:S1Fusion}.  As usual, the computation splits into an oscillator part and a zero mode part.

\subsubsection{Oscillator contribution}
\label{appsubsec:OscillatorFusion}

For the oscillators, what we must compute is the set of quantities
\begin{multline}
\mathcal{C}_n=\vphantom{\langle}_2\langle 0|e^{S^{(\e)}_{11}a_n^{1\,\dagger}\wta_n^{1\,\dagger}-S^{(\e)}_{12}a_n^{1\,\dagger}a_n^2-S^{(\e)}_{21}\wta_n^2\wta_n^{1\,\dagger}+S^{(\e)}_{22}\wta_n^2a_n^2}\\
\times q^{n(a_n^2a_n^{2\,\dagger}+\wta_n^2\wta_n^{2\,\dagger})}e^{S^{(\e')}_{11}a_n^{2\,\dagger}\wta_n^{2\,\dagger}-S^{(\e')}_{12}a_n^{2\,\dagger}a_n^3-S^{(\e')}_{21}\wta_n^3\wta_n^{2\,\dagger}+S^{(\e')}_{22}\wta_n^3a_n^3}|0\rangle_2,
\end{multline}
where $\e$ and $\e'$ are $\pm 1$.  Here we may view the oscillators $a_n^{1\,\dagger}$, $\wta_n^{1\,\dagger}$, $a_n^3$, and $\wta_n^3$ as c-numbers for the purposes of this computation.  We can use the same logic as in (\ref{eq:Absorbq}) in order to absorb the $q$-dependence into the second exponential, and write
\be
\mathcal{C}_n=\vphantom{\langle}_2\langle 0|e^Ae^B|0\rangle_2,
\ee
where
\bea
A &=& S^{(\e)}_{11}a_n^{1\,\dagger}\wta_n^{1\,\dagger}-S^{(\e)}_{12}a_n^{1\,\dagger}a_n^2-S^{(\e)}_{21}\wta_n^2\wta_n^{1\,\dagger}+S^{(\e)}_{22}\wta_n^2a_n^2,\\
B &=& q^{2n}S^{(\e')}_{11}a_n^{2\,\dagger}\wta_n^{2\,\dagger}-q^nS^{(\e')}_{12}a_n^{2\,\dagger}a_n^3-q^nS^{(\e')}_{21}\wta_n^3\wta_n^{2\,\dagger}+S^{(\e')}_{22}\wta_n^3a_n^3.
\eea

To compute $\mathcal{C}_n$, we will show that it satisfies a simple system of first order differential equations.  Indeed, we have
\bea
\frac{\p\mathcal{C}_n}{\p a_n^{1\,\dagger}} &=& \vphantom{\langle}_2\langle 0|e^A\lp S^{(\e)}_{11}\wta_n^{1\,\dagger}-S^{(\e)}_{12}a_n^2\rp e^B|0\rangle_2\non\\
&=& S^{(\e)}_{11}\wta_n^{1\,\dagger}\mathcal{C}_n-S^{(\e)}_{12}\vphantom{\langle}_2\langle 0|e^Ae^B\lp q^{2n}S^{(\e')}_{11}\wta_n^{2\,\dagger}-q^nS^{(\e')}_{12}a_n^3\rp|0\rangle_2.
\eea
Meanwhile,
\be
\frac{\p\mathcal{C}_n}{\p\wta_n^3}=\vphantom{\langle}_2\langle 0|e^Ae^B\lp -q^nS^{(\e')}_{21}\wta_n^{2\,\dagger}+S^{(\e')}_{22}a_n^3\rp|0\rangle_2.
\ee
Hence,
\be
\frac{\p\mathcal{C}_n}{\p a_n^{1\,\dagger}}=\lp S^{(\e)}_{11}\wta_n^{1\,\dagger}+\e'q^n\frac{S^{(\e)}_{12}}{S^{(\e')}_{21}}a_n^3\rp\mathcal{C}_n+q^n\frac{S^{(\e)}_{12}S^{(\e')}_{11}}{S^{(\e')}_{21}}\frac{\p\mathcal{C}_n}{\p\wta_n^3},
\ee
where we used the fact that $\det(S^{(\e')})=-\e'$.  Similar manipulations lead to
\bea
\frac{\p\mathcal{C}_n}{\p\wta_n^{1\,\dagger}} &=& \lp S^{(\e)}_{11}a_n^{1\,\dagger}+\e'q^n\frac{S^{(\e)}_{21}}{S^{(\e')}_{12}}\wta_n^3\rp\mathcal{C}_n+q^n\frac{S^{(\e)}_{21}S^{(\e')}_{11}}{S^{(\e')}_{12}}\frac{\p\mathcal{C}_n}{\p a_n^3},\\
\frac{\p\mathcal{C}_n}{\p a_n^3} &=& \lp S^{(\e')}_{22}\wta_n^3+\e q^n\frac{S^{(\e')}_{12}}{S^{(\e)}_{21}}a_n^{1\,\dagger}\rp\mathcal{C}_n+q^n\frac{S^{(\e)}_{22}S^{(\e')}_{12}}{S^{(\e)}_{21}}\frac{\p\mathcal{C}_n}{\p\wta_n^{1\,\dagger}},\\
\frac{\p\mathcal{C}_n}{\p\wta_n^3} &=& \lp S^{(\e')}_{22}a_n^3+\e q^n\frac{S^{(\e')}_{21}}{S^{(\e)}_{12}}\wta_n^{1\,\dagger}\rp\mathcal{C}_n+q^n\frac{S^{(\e)}_{22}S^{(\e')}_{21}}{S^{(\e)}_{12}}\frac{\p\mathcal{C}_n}{\p a_n^{1\,\dagger}}.
\eea
Then a little algebra gives
\bea
\frac{\p\mathcal{C}_n}{\p a_n^{1\,\dagger}} &=& \lp 1-q^{2n}S^{(\e)}_{22}S^{(\e')}_{11}\rp^{-1}\ls\lp S^{(\e)}_{11}+\e q^{2n}S^{(\e')}_{11}\rp\wta_n^{1\,\dagger}+q^nS^{(\e)}_{12}S^{(\e')}_{12}a_n^3\rs\mathcal{C}_n,\\
\frac{\p\mathcal{C}_n}{\p\wta_n^{1\,\dagger}} &=& \lp 1-q^{2n}S^{(\e)}_{22}S^{(\e')}_{11}\rp^{-1}\ls\lp S^{(\e)}_{11}+\e q^{2n}S^{(\e')}_{11}\rp a_n^{1\,\dagger}+q^nS^{(\e)}_{21}S^{(\e')}_{21}\wta_n^3\rs\mathcal{C}_n,\\
\frac{\p\mathcal{C}_n}{\p a_n^3} &=& \lp 1-q^{2n}S^{(\e)}_{22}S^{(\e')}_{11}\rp^{-1}\ls q^nS^{(\e)}_{12}S^{(\e')}_{12}a_n^{1\,\dagger}+\lp S^{(\e')}_{22}+\e' q^{2n}S^{(\e)}_{22}\rp\wta_n^3\rs\mathcal{C}_n,\\
\frac{\p\mathcal{C}_n}{\p\wta_n^3} &=& \lp 1-q^{2n}S^{(\e)}_{22}S^{(\e')}_{11}\rp^{-1}\ls q^nS^{(\e)}_{21}S^{(\e')}_{21}\wta_n^{1\,\dagger}+\lp S^{(\e')}_{22}+\e'q^{2n}S^{(\e)}_{22}\rp a_n^3\rs\mathcal{C}_n.
\eea
This fixes the form of $\mathcal{C}_n$ up to an arbitrary function of $q$,
\begin{multline}
\mathcal{C}_n=f(q)\exp\ls\lp 1-q^{2n}S^{(\e)}_{22}S^{(\e')}_{11}\rp^{-1}\lp\lp S^{(\e)}_{11}+\e q^{2n}S^{(\e')}_{11}\rp a_n^{1\,\dagger}\wta_n^{1\,\dagger}+q^nS^{(\e)}_{12}S^{(\e')}_{12}a_n^{1\,\dagger}a_n^3\right.\right.\\
\left.\left. +q^nS^{(\e)}_{21}S^{(\e')}_{21}\wta_n^3\wta_n^{1\,\dagger}+\lp S^{(\e')}_{22}+\e'q^{2n}S^{(\e)}_{22}\rp\wta_n^3a_n^3\rp\rs.
\end{multline}
Finally, we can fix $f(q)$ by formally evaluating $\mathcal{C}_n$ with $a_n^{1\,\dagger}=\wta_n^{1\,\dagger}=a_n^3=\wta_n^3=0$.  We find
\be
f(q)=\vphantom{\langle}_2\langle 0|e^{S^{(\e)}_{22}\wta_n^2a_n^2}e^{q^{2n}S^{(\e')}_{11}a_n^{2\,\dagger}\wta_n^{2\,\dagger}}|0\rangle_2.
\ee
Writing
\be
\vphantom{\langle}_2\langle 0|\lp\wta_n^2a_n^2\rp^k\lp a_n^{2\,\dagger}\wta_n^{2\,\dagger}\rp^\ell|0\rangle_2=\d_{k\ell}c_k,
\ee
we have $c_0=1$ and can show $c_k=k^2c_{k-1}$, which implies by induction that $c_k=(k!)^2$, and hence
\be
f(q)=\sum_{k,\ell=0}^\infty\frac{\lp S^{(\e)}_{22}\rp^k\lp q^{2n}S^{(\e')}_{11}\rp^\ell}{k!\ell!}\d_{k\ell}c_k=\frac{1}{1-q^{2n}S^{(\e)}_{22}S^{(\e')}_{11}}.
\ee
Thus,
\be
\mathcal{C}_n=\frac{1}{1-q^{2n}S^{(\e)}_{22}S^{(\e')}_{11}}e^{M^{(\e\e')}_{11}a_n^{1\,\dagger}\wta_n^{1\,\dagger}-M^{(\e\e')}_{12}a_n^{1\,\dagger}a_n^3-M^{(\e\e')}_{21}\wta_n^3\wta_n^{1\,\dagger}+M^{(\e\e')}_{22}\wta_n^3a_n^3},
\ee
where
\be
M^{(\e\e')}=\frac{1}{1-q^{2n}S^{(\e)}_{22}S^{(\e')}_{11}}\lp\begin{matrix}S^{(\e)}_{11}+\e q^{2n}S^{(\e')}_{11} & -q^nS^{(\e)}_{12}S^{(\e')}_{12} \\ -q^nS^{(\e)}_{21}S^{(\e')}_{21} & S^{(\e')}_{22}+\e'q^{2n}S^{(\e)}_{22}\end{matrix}\rp.
\ee

We can obtain the special case where the second interface is factorized by formally setting $a_n^3=\wta_n^3=0$, while taking $S^{(\e')}_{11}=-\e'$, $S^{(\e)}_{11}=-\e\cos(2\theta)$, $S^{(\e)}_{22}=\cos(2\theta)$, giving a result
\be
\mathcal{C}_n=\frac{1}{1+\e'q^{2n}\cos(2\theta)}e^{-\e\e'a_n^{1\,\dagger}\wta_n^{1\,\dagger}}.
\ee

\subsubsection{Zero mode contribution}

The zero mode piece depends on whether each interface is $\mcO^{(+)}$ or $\mcO^{(-)}$.  For the product of two $\mcO^{(+)}$ interfaces, we have
\begin{multline}
\mathcal{Z}=\lp\sum_{N,M\in\Z}e^{-iM\al-iN\beta}|-k_2N,k_1M\rangle_1\langle -k_1N,k_2M|_2\rp\\
\times\lp\sum_{N',M'\in\Z}e^{-iM'\al'-iN'\beta'}q^{\lp\frac{k_2'N'}{2R_2}\rp^2+(k_1'M'R_2)^2}|-k_2'N',k_1'M'\rangle_2\langle -k_1'N',k_2'M'|_3\rp
\end{multline}
Define
\be
\label{eq:BigGDef}
G_{ab'}=\gcd(k_a,k_b').
\ee
Then the theory 2 matrix element will vanish unless
\be
N=\frac{k_2'N''}{G_{12'}},\quad N'=\frac{k_1N''}{G_{12'}},\quad M=\frac{k_1'M''}{G_{21'}},\quad M'=\frac{k_2M''}{G_{21'}},
\ee
for some arbitrary integers $N''$ and $M''$.  Then we get (relabeling $N''\rr N$, $M''\rr M$),
\begin{multline}
\mathcal{Z}=\sum_{N,M\in\Z}e^{-iM\frac{k_1'\al+k_2\al'}{G_{21'}}-iN\frac{k_2'\beta+k_1\beta'}{G_{12'}}}q^{\lp\frac{k_1k_2'N}{2G_{12'}R_2}\rp^2+\lp\frac{k_2k_1'MR_2}{G_{21'}}\rp^2}\\
\times|-\frac{k_2k_2'N}{G_{12'}},\frac{k_1k_1'M}{G_{21'}}\rangle_1\langle -\frac{k_1k_1'N}{G_{12'}},\frac{k_2k_2'M}{G_{21'}}|_3.
\end{multline}

\subsubsection{Regularization}

We'll again start by considering only $\mcO^{(+)}$ interfaces.  Note that $\mathcal{Z}$ has a smooth limit as $q\rr 1$, and we also have
\be
\lim_{q\rr 1}M^{(++)}=S^{(+)}(\Theta),
\ee
where
\be
\tan\Theta=\tan\vartheta\tan\vartheta'=\frac{k_2R_2}{k_1R_1}\frac{k_2'R_3}{k_1'R_2}=\frac{k_2k_2'R_3}{k_1k_1'R_1}.
\ee

The only place we have a potential divergence is from the product of the $f(q)$, which contributes (we do the calculation for arbitrary $\e$ and $\e'$ for use below)
\bea
\prod_{n=1}^\infty\frac{1}{1-q^{2n}S^{(\e)}_{22}S^{(\e')}_{11}} &=& \exp\ls -\sum_{n=1}^\infty\ln\lp1-q^{2n}S^{(\e)}_{22}S^{(\e')}_{11}\rp\rs\non\\
&=& \exp\ls\sum_{n=1}^\infty\sum_{k=1}^\infty\frac{\lp q^{2n}S^{(\e)}_{22}S^{(\e')}_{11}\rp^k}{k}\rs\non\\
&=& \exp\ls\sum_{k=1}^\infty\frac{\lp S^{(\e)}_{22}S^{(\e')}_{11}\rp^k}{k}\frac{q^{2k}}{1-q^{2k}}\rs.
\eea
Now setting $q=e^{-t}$ and expanding in small $t$, we get
\be
\sum_{k=1}^\infty\frac{\lp S^{(\e)}_{22}S^{(\e')}_{11}\rp^k}{k}\lp\frac{1}{2kt}-\hlf+\mcO(t)\rp=\frac{1}{2t}\operatorname{Li}_2(S^{(\e)}_{22}S^{(\e')}_{11})+\hlf\ln(1-S^{(\e)}_{22}S^{(\e')}_{11})+\mcO(t),
\ee
where $\operatorname{Li}_2(x)=\sum_{k=1}^\infty x^k/k^2$ is the dilogarithm function.  Thus to regulate we take
\be
d=-\frac{1}{4\pi}\operatorname{Li}_2(S^{(\e)}_{22}S^{(\e')}_{11}).
\ee
Note that if either interface is topological, in which case either $S^{(\e)}$ or $S^{(\e')}$ is purely off-diagonal, then $d=0$ and we don't actually need a regulator.

With $d$ computed, we can take $t$ to zero, obtaining
\be
\lim_{t\rr 0}e^{2\pi dt}\prod_{n=1}^\infty\frac{1}{1-q^{2n}S^{(\e)}_{22}S^{(\e')}_{11}}=\sqrt{1-S^{(\e)}_{22}S^{(\e')}_{11}}.
\ee
For $\e=\e'=+$, this is
\be
\sqrt{1-S^{(+)}_{22}S^{(+)}_{11}}=\sqrt{1+\cos(2\vartheta)\cos(2\vartheta')}=\sqrt{\frac{2R_2^2\lp (k_1k_1'R_1)^2+(k_2k_2'R_3)^2\rp}{\lp (k_1R_1)^2+(k_2R_2)^2\rp\lp (k_1'R_2)^2+(k_2'R_3)^2\rp}}.
\ee

\subsubsection{Full result}

Returning to the full fusion of two ND interfaces, we find
\begin{multline}
\mcO^{(+)}_{k_1,k_2}(\al;\beta)\circ\mcO^{(+)}_{k_1',k_2'}(\al';\beta')\\
=\sqrt{\frac{(k_1k_1'R_1)^2+(k_2k_2'R_3)^2}{2R_1R_3}}\lp\prod_{n=1}^\infty e^{S^{(+)}_{11}a_n^{1\,\dagger}\wta_n^{1\,\dagger}-S^{(+)}_{12}a_n^{1\,\dagger}a_n^3-S^{(+)}_{21}\wta_n^3\wta_n^{1\,\dagger}+S^{(+)}_{22}\wta_n^3a_n^3}\rp\\
\times\sum_{N,M\in\Z}e^{-iM\frac{k_1'\al+k_2\al'}{G_{21'}}-iN\frac{k_2'\beta+k_1\beta'}{G_{12'}}}|-\frac{k_2k_2'N}{G_{12'}},\frac{k_1k_1'M}{G_{21'}}\rangle_1\langle -\frac{k_1k_1'N}{G_{12'}},\frac{k_2k_2'M}{G_{21'}}|_3,
\end{multline}
where $G_{12'}$ and $G_{21'}$ are defined by (\ref{eq:BigGDef}).

Unless $G_{12'}=G_{21'}$ (in which case $G_{12'}=G_{21'}=1$, since one can check that $G_{12'}$ and $G_{21'}$ are relatively prime), this is not yet manifestly in the desired form of being a linear combination of our defects, since for instance the coefficient of $-N$ in the ket does not agree with the coefficient of $M$ in the bra.  To fix this, we note that we can trivially rewrite the sums by replacing $N=N'/G_{21'}$, $M=M'/G_{12'}$ and summing over $N'$ and $M'$, but restricted to multiples of $G_{21'}$ and $G_{12'}$ respectively.  Next we observe that this restriction can be implemented by summing over all integers while inserting some finite sum projectors.  Explicitly, for any summand $c_{N'M'}$ we have
\be
\sum_{N',M'\in\Z,G_{21'}|N',G_{12'}|M'}c_{N'M'}=\frac{1}{G_{12'}G_{21'}}\sum_{j=0}^{G_{21'}-1}\sum_{j'=0}^{G_{12'}-1}\sum_{N,M\in\Z}e^{-2\pi i\frac{Nj}{G_{21'}}-2\pi i\frac{Mj'}{G_{12'}}}c_{NM}.
\ee
Using this trick, we have
\begin{multline}
\mcO^{(+)}_{k_1,k_2}(\al;\beta)\circ\mcO^{(+)}_{k_1',k_2'}(\al';\beta')=\sqrt{\frac{K_1^2R_1^2+K_2^2R_3^2}{2R_1R_3}}\lp\prod_{n=1}^\infty e^{S^{(+)}_{11}a_n^{1\,\dagger}\wta_n^{1\,\dagger}-S^{(+)}_{12}a_n^{1\,\dagger}a_n^3-S^{(+)}_{21}\wta_n^3\wta_n^{1\,\dagger}+S^{(+)}_{22}\wta_n^3a_n^3}\rp\\
\times\sum_{j=0}^{G_{21'}-1}\sum_{j'=0}^{G_{12'}-1}\sum_{N,M\in\Z}e^{-iM\lp\al''+\frac{2\pi j'}{G_{12'}}\rp-iN\lp\beta''+\frac{2\pi j}{G_{21'}}\rp}|-K_2N,K_1M\rangle_1\langle -K_1N,K_2M|_3,
\end{multline}
where
\be
K_1=\frac{k_1k_1'}{G_{12'}G_{21'}},\qquad K_2=\frac{k_2k_2'}{G_{12'}G_{21'}}.
\ee
So finally, we can write the general fusion of two ND defects as
\be
\mcO^{(+)}_{k_1,k_2}(\al;\beta)\circ\mcO^{(+)}_{k_1',k_2'}(\al';\beta')=\sum_{j=0}^{G_{21'}-1}\sum_{j'=0}^{G_{12'}-1}\mcO^{(+)}_{K_1,K_2}(\al''+\frac{2\pi j'}{G_{12'}};\beta''+\frac{2\pi j}{G_{21'}}).
\ee

Similar manipulations can be used to compute the other fusions,
\bea
\mcO^{(+)}_{k_1,k_2}(\al;\beta)\circ\mcO^{(-)}_{k_1',k_2'}(\al';\beta') &=& \sum_{j=0}^{G_{11'}-1}\sum_{j'=0}^{G_{22'}-1}\mcO^{(-)}_{K_1',K_2'}(\al'''+\frac{2\pi j'}{G_{22'}};\beta'''+\frac{2\pi j}{G_{11'}}),\\
\mcO^{(-)}_{k_1,k_2}(\al;\beta)\circ\mcO^{(+)}_{k_1',k_2'}(\al';\beta') &=& \sum_{j=0}^{G_{21'}-1}\sum_{j'=0}^{G_{12'}-1}\mcO^{(-)}_{K_1,K_2}(\al''+\frac{2\pi j'}{G_{12'}};\beta''+\frac{2\pi j}{G_{21'}}),\\
\mcO^{(-)}_{k_1,k_2}(\al;\beta)\circ\mcO^{(-)}_{k_1',k_2'}(\al';\beta') &=& \sum_{j=0}^{G_{11'}-1}\sum_{j'=0}^{G_{22'}-1}\mcO^{(+)}_{K_1',K_2'}(\al'''+\frac{2\pi j'}{G_{22'}};\beta'''+\frac{2\pi j}{G_{11'}}),
\eea
where we have defined
\be
\al''=\frac{k_1'\al+k_2\al'}{G_{12'}G_{21'}},\qquad\beta''=\frac{k_2'\beta+k_1\beta'}{G_{12'}G_{21'}},
\ee
as well as
\be
K_1'=\frac{k_2k_1'}{G_{11'}G_{22'}},\quad K_2'=\frac{k_1k_2'}{G_{11'}G_{22'}},\quad\al'''=\frac{k_1\al'-k_1'\beta}{G_{11'}G_{22'}},\quad\beta'''=\frac{k_2\beta'-k_2'\al}{G_{11'}G_{22'}}.
\ee

These results can be combined to give (\ref{eq:S1Fusion}).

\subsection{Twisted sector fusion}

For the pieces of interfaces involving twisted sector states, the ground state part is very simple.  We simply need to consider the contribution from the oscillator portion, and in particular implement the regularization procedure.  In the computation of $\mathcal{C}_n$ in section \ref{appsubsec:OscillatorFusion}, we never used the fact that $n$, was an integer, so repeating the computation for half-integer $r$ gives, the same result,
\be
\mathcal{C}_r=\frac{1}{1-q^{2r}S^{(\e)}_{22}S^{(\e')}_{11}}e^{M^{(\e\e')}_{11}a_r^{1\,t\,\dagger}\wta_r^{1\,t\,\dagger}-M^{(\e\e')}_{12}a_r^{1\,t\,\dagger}a_r^{3\,t}-M^{(\e\e')}_{21}\wta_r^{3\,t}\wta_r^{1\,t\,\dagger}+M^{(\e\e')}_{22}\wta_r^{3\,t}a_r^{3\,t}},
\ee
where
\be
M^{(\e\e')}=\frac{1}{1-q^{2r}S^{(\e)}_{22}S^{(\e')}_{11}}\lp\begin{matrix}S^{(\e)}_{11}+\e q^{2r}S^{(\e')}_{11} & -q^rS^{(\e)}_{12}S^{(\e')}_{12} \\ -q^rS^{(\e)}_{21}S^{(\e')}_{21} & S^{(\e')}_{22}+\e'q^{2r}S^{(\e)}_{22}\end{matrix}\rp.
\ee

For the product of a non-factorized interface with a boundary state, we get the result by setting $a_r^{3\,t}=\wta_r^{3\,t}=0$,
\be
\mathcal{C}_r=\frac{1}{1+\e'q^{2r}\cos(2\theta)}e^{-\e\e'a_r^{1\,t\,\dagger}\wta_r^{1\,t\,\dagger}}.
\ee

For the reqularization, the potential divergences again come only from the product of the factors in front of $\mathcal{C}_r$,
\bea
\prod_{r\in\Z+\hlf,r>0}\frac{1}{1-q^{2r}S^{(\e)}_{22}S^{(\e')}_{11}} &=& \exp\ls -\sum_{r\in\Z+\hlf,r>0}\ln\lp1-q^{2r}S^{(\e)}_{22}S^{(\e')}_{11}\rp\rs\non\\
&=& \exp\ls\sum_{r\in\Z+\hlf,r>0}\sum_{k=1}^\infty\frac{\lp q^{2r}S^{(\e)}_{22}S^{(\e')}_{11}\rp^k}{k}\rs\non\\
&=& \exp\ls\sum_{k=1}^\infty\frac{\lp S^{(\e)}_{22}S^{(\e')}_{11}\rp^k}{k}\lp\frac{q^k}{1-q^k}-\frac{q^{2k}}{1-q^{2k}}\rp\rs\non\\
&=& \exp\ls\sum_{k=1}^\infty\frac{\lp S^{(\e)}_{22}S^{(\e')}_{11}\rp^k}{k}\frac{q^k}{1-q^{2k}}\rs.
\eea
Now setting $q=e^{-t}$ and expanding in small $t$, we get in the exponent
\be
\sum_{k=1}^\infty\frac{\lp S^{(\e)}_{22}S^{(\e')}_{11}\rp^k}{k}\lp\frac{1}{2kt}+\mcO(t)\rp=\frac{1}{2t}\operatorname{Li}_2(S^{(\e)}_{22}S^{(\e')}_{11})+\mcO(t),
\ee
where $\operatorname{Li}_2(x)=\sum_{k=1}^\infty x^k/k^2$ is the dilogarithm function.  Thus to regulate we take
\be
d=-\frac{1}{4\pi}\operatorname{Li}_2(S^{(\e)}_{22}S^{(\e')}_{11}).
\ee
This is the same value of $d$ that we got in the untwisted sector, indicating that both sectors can contribute to the fusion, as one might expect.  The difference here is that there is no extra normalization factor, which also matches our expectations.  Indeed, taking $t$ to zero, we get
\be
\lim_{t\rr 0}e^{2\pi dt}\prod_{r\in\Z+\hlf,r>0}\frac{1}{1-q^{2r}S^{(\e)}_{22}S^{(\e')}_{11}}=1.
\ee

\subsection{Non-factorized with factorized fusion}
\label{appsubsec:ZMFusionSpecial}

We can use the same techniques to compute the fusion of one non-factorized $S^1|S^1$ interface, $\mcO^{(\pm)}_{k_1,k_2}(\al;\beta)$, $k_1,k_2\ne 0$, with an $S^1$ boundary state.  For example, we have
\bea
\mcO^{(+)}_{k_1,k_2}(\al;\beta)\circ\lp|N(\al')\rangle\rangle_2\langle\langle B'|_3\rp &=& k_1\sqrt{R_1}\lp\prod_{n=1}^\infty e^{-a_n^{1\,\dagger}\wta_n^{1\,\dagger}}\rp\sum_{N,M,M'\in\Z}e^{-iM\al-iN\beta-iM'\al'}\non\\
&& \quad\times|-k_2N,k_1M\rangle_1\langle -k_1N,k_2M|0,M'\rangle_2\langle\langle B'|_3\non\\
&=& k_1\sqrt{R_1}\lp\prod_{n=1}^\infty e^{-a_n^{1\,\dagger}\wta_n^{1\,\dagger}}\rp\sum_{M\in\Z}e^{-iM(\al+k_2\al')}|0,k_1M\rangle_1\langle\langle B'|_3\non\\
&=& \sum_{j=0}^{k_1-1}|N(\frac{\al+k_2\al'+2\pi j}{k_1})\rangle\rangle_1\langle\langle B'|_3.
\eea
Similarly, we can compute
\bea
\mcO^{(+)}_{k_1,k_2}(\al;\beta)\circ\lp|D(\beta')\rangle\rangle_2\langle\langle B'|_3\rp &=& \sum_{j=0}^{k_2-1}|D(\frac{-\beta+k_1\beta'+2\pi j}{k_2})\rangle\rangle_1\langle\langle B'|_3,\\
\mcO^{(-)}_{k_1,k_2}(\al;\beta)\circ\lp|N(\al')\rangle\rangle_2\langle\langle B'|_3\rp &=& \sum_{j=0}^{k_1-1}|D(\frac{\al+k_2\al'+2\pi j}{k_1})\rangle\rangle_1\langle\langle B'|_3,\\
\mcO^{(-)}_{k_1,k_2}(\al;\beta)\circ\lp|D(\beta')\rangle\rangle_2\langle\langle B'|_3\rp &=& \sum_{j=0}^{k_2-1}|N(\frac{-\beta+k_1\beta'+2\pi j}{k_2})\rangle\rangle_1\langle\langle B'|_3.
\eea

For fusions of this sort that also involve twisted sectors, we also need combinations of terms like
\begin{multline}
\lp\lp\prod_{r\in\Z+\hlf,r>0}e^{S^{(\eta)}_{11}a_r^{1\,t\,\dagger}\wta_r^{1\,t\,\dagger}-S^{(\eta)}_{12}a_r^{1\,t\,\dagger}a_r^{2\,t}-S^{(\eta)}_{21}\wta_r^{2\,t}\wta_r^{1\,t\,\dagger}+S^{(\eta)}_{22}\wta_r^{2\,t}a_r^{2\,t}}\rp|\g_0\rangle_1\langle\d_0|_2\rp\\
\circ\lp\lp\prod_{r\in\Z+\hlf,r>0}e^{-\eta'a_r^{2\,t\,\dagger}\wta_r^{2\,t\,\dagger}}\rp|\g'_0\rangle_2\langle\langle B'|_3\rp\\
=\d_{\d_0\g'_0}\lp\prod_{r\in\Z+\hlf,r>0}e^{-\eta\eta'a_r^{1\,t\,\dagger}\wta_r^{1\,t\,\dagger}}\rp|\g_0\rangle_1\langle\langle B'|_3.
\end{multline}

\bibliographystyle{utphys}


\begin{thebibliography}{10}

\bibitem{cardy}
J.~Cardy, ``{Conformal symmetry and critical surface behavior},''
{\em Nucl. Phys.} {\bf B240} (1984)  .

\bibitem{Cardy:1989ir}
J.~L. Cardy, ``{Boundary Conditions, Fusion Rules and the Verlinde Formula},''
\href{http://dx.doi.org/10.1016/0550-3213(89)90521-X}{{\em Nucl. Phys.} {\bf
  B324} (1989)  581--596}.

\bibitem{Cardy:1991tv}
J.~L. Cardy and D.~C. Lewellen, ``{Bulk and boundary operators in conformal
  field theory},''
\href{http://dx.doi.org/10.1016/0370-2693(91)90828-E}{{\em Phys. Lett.} {\bf
  B259} (1991)  274--278}.

\bibitem{Lewellen:1991tb}
D.~C. Lewellen, ``{Sewing constraints for conformal field theories on surfaces
  with boundaries},''
\href{http://dx.doi.org/10.1016/0550-3213(92)90370-Q}{{\em Nucl. Phys.} {\bf
  B372} (1992)  654--682}.

\bibitem{Recknagel:1998ih}
A.~Recknagel and V.~Schomerus, ``{Boundary deformation theory and moduli spaces
  of D-branes},'' \href{http://dx.doi.org/10.1016/S0550-3213(99)00060-7}{{\em
  Nucl. Phys.} {\bf B545} (1999)  233--282},
\href{http://arxiv.org/abs/hep-th/9811237}{{\tt arXiv:hep-th/9811237
  [hep-th]}}.

\bibitem{Brunner:1999}
I.~Brunner, R.~Entin, and C.~Romelsberger, ``{D-branes on T**4 / Z(2) and T
  duality},'' \href{http://dx.doi.org/10.1088/1126-6708/1999/06/016}{{\em JHEP}
  {\bf 06} (1999)  016},
\href{http://arxiv.org/abs/hep-th/9905078}{{\tt arXiv:hep-th/9905078
  [hep-th]}}.

\bibitem{Gaberdiel:2001zq}
M.~R. Gaberdiel and A.~Recknagel, ``{Conformal boundary states for free bosons
  and fermions},'' \href{http://dx.doi.org/10.1088/1126-6708/2001/11/016}{{\em
  JHEP} {\bf 11} (2001)  016},
\href{http://arxiv.org/abs/hep-th/0108238}{{\tt arXiv:hep-th/0108238
  [hep-th]}}.

\bibitem{Gaberdiel:2008rk}
M.~R. Gaberdiel and O.~Schlotterer, ``{Bulk induced boundary perturbations for
  N=1 superconformal field theories},''
  \href{http://dx.doi.org/10.1088/1751-8113/42/11/115209}{{\em J. Phys.} {\bf
  A42} (2009)  115209},
\href{http://arxiv.org/abs/0810.4719}{{\tt arXiv:0810.4719 [hep-th]}}.

\bibitem{bachas02}
C.~Bachas, J.~de~Boer, R.~Dijkgraaf, and H.~Ooguri, ``{Permeable conformal
  walls and holography},''
  \href{http://dx.doi.org/10.1088/1126-6708/2002/06/027}{{\em JHEP} {\bf 06}
  (2002)  027},
\href{http://arxiv.org/abs/hep-th/0111210}{{\tt arXiv:hep-th/0111210
  [hep-th]}}.

\bibitem{bachas07}
C.~Bachas and I.~Brunner, ``{Fusion of conformal interfaces},''
  \href{http://dx.doi.org/10.1088/1126-6708/2008/02/085}{{\em JHEP} {\bf 02}
  (2008)  085},
\href{http://arxiv.org/abs/0712.0076}{{\tt arXiv:0712.0076 [hep-th]}}.

\bibitem{fuchs07}
J.~Fuchs, M.~R. Gaberdiel, I.~Runkel, and C.~Schweigert, ``{Topological defects
  for the free boson CFT},''
  \href{http://dx.doi.org/10.1088/1751-8113/40/37/016}{{\em J. Phys.} {\bf A40}
  (2007)  11403},
\href{http://arxiv.org/abs/0705.3129}{{\tt arXiv:0705.3129 [hep-th]}}.

\bibitem{gaiotto12}
D.~Gaiotto, ``{Domain Walls for Two-Dimensional Renormalization Group Flows},''
  \href{http://dx.doi.org/10.1007/JHEP12(2012)103}{{\em JHEP} {\bf 12} (2012)
  103},
\href{http://arxiv.org/abs/1201.0767}{{\tt arXiv:1201.0767 [hep-th]}}.

\bibitem{Quella:2002CT}
T.~Quella and V.~Schomerus, ``{Symmetry breaking boundary states and defect
  lines},'' \href{http://dx.doi.org/10.1088/1126-6708/2002/06/028}{{\em JHEP}
  {\bf 06} (2002)  028},
\href{http://arxiv.org/abs/hep-th/0203161}{{\tt arXiv:hep-th/0203161
  [hep-th]}}.

\bibitem{brunner03}
I.~Brunner, M.~Herbst, W.~Lerche, and B.~Scheuner, ``{Landau-Ginzburg
  realization of open string TFT},''
  \href{http://dx.doi.org/10.1088/1126-6708/2006/11/043}{{\em JHEP} {\bf 11}
  (2006)  043},
\href{http://arxiv.org/abs/hep-th/0305133}{{\tt arXiv:hep-th/0305133
  [hep-th]}}.

\bibitem{brunner07}
I.~Brunner and D.~Roggenkamp, ``{B-type defects in Landau-Ginzburg models},''
  \href{http://dx.doi.org/10.1088/1126-6708/2007/08/093}{{\em JHEP} {\bf 0708}
  (2007)  093},
\href{http://arxiv.org/abs/0707.0922}{{\tt arXiv:0707.0922 [hep-th]}}.

\bibitem{Konechny:2015qla}
A.~Konechny, ``{Fusion of conformal interfaces and bulk induced boundary RG
  flows},'' \href{http://dx.doi.org/10.1007/JHEP12(2015)114}{{\em JHEP} {\bf
  12} (2015)  114},
\href{http://arxiv.org/abs/1509.07787}{{\tt arXiv:1509.07787 [hep-th]}}.

\bibitem{Graham:2003nc}
K.~Graham and G.~M.~T. Watts, ``{Defect lines and boundary flows},''
  \href{http://dx.doi.org/10.1088/1126-6708/2004/04/019}{{\em JHEP} {\bf 04}
  (2004)  019},
\href{http://arxiv.org/abs/hep-th/0306167}{{\tt arXiv:hep-th/0306167
  [hep-th]}}.

\bibitem{Fuchs:2015ska}
J.~Fuchs and C.~Schweigert, ``{Surface defects and symmetries},''
\href{http://dx.doi.org/10.1088/1742-6596/597/1/012002}{{\em J. Phys. Conf.
  Ser.} {\bf 597} (2015) no.~1, 012002}.

\bibitem{Bachas:2012bj} 
  C.~Bachas, I.~Brunner and D.~Roggenkamp,
  JHEP {\bf 1210}, 039 (2012)
  doi:10.1007/JHEP10(2012)039
  [arXiv:1205.4647 [hep-th]].
  
\bibitem{affleck}
M.~Oshikawa and I.~Affleck, ``{Boundary conformal field theory approach to the
  critical two-dimensional Ising model with a defect line},''
  \href{http://dx.doi.org/10.1016/S0550-3213(97)00219-8}{{\em Nucl. Phys.} {\bf
  B495} (1997)  533--582},
\href{http://arxiv.org/abs/cond-mat/9612187}{{\tt arXiv:cond-mat/9612187
  [cond-mat]}}.

\bibitem{brunner07a}
I.~Brunner and D.~Roggenkamp, ``{Defects and bulk perturbations of boundary
  Landau-Ginzburg orbifolds},''
  \href{http://dx.doi.org/10.1088/1126-6708/2008/04/001}{{\em JHEP} {\bf 04}
  (2008)  001},
\href{http://arxiv.org/abs/0712.0188}{{\tt arXiv:0712.0188 [hep-th]}}.

\bibitem{Becker:2017}
M.~Becker, Y.~Cabrera, and D.~Robbins, ``{Defects and boundary RG flows in $
  \mathbb{C}/{\mathbb{Z}}_d $},''
  \href{http://dx.doi.org/10.1007/JHEP02(2017)007}{{\em JHEP} {\bf 02} (2017)
  007},
\href{http://arxiv.org/abs/1611.01133}{{\tt arXiv:1611.01133 [hep-th]}}.

\bibitem{Frohlich09gb}
J.~Frohlich, J.~Fuchs, I.~Runkel, and C.~Schweigert, ``{Defect lines,
  dualities, and generalised orbifolds},'' in {\em {Proceedings, 16th
  International Congress on Mathematical Physics (ICMP09): Prague, Czech
  Republic, August 3-8, 2009}}.
\newblock 2009.
\newblock \href{http://arxiv.org/abs/0909.5013}{{\tt arXiv:0909.5013
  [math-ph]}}.
\newblock
\url{https://inspirehep.net/record/832556/files/arXiv:0909.5013.pdf}.
\newblock

\bibitem{Elitzur:2013ut}
S.~Elitzur, B.~Karni, E.~Rabinovici, and G.~Sarkissian, ``{Defects,
  Super-Poincar{\'e} line bundle and Fermionic T-duality},''
  \href{http://dx.doi.org/10.1007/JHEP04(2013)088}{{\em JHEP} {\bf 04} (2013)
  088},
\href{http://arxiv.org/abs/1301.6639}{{\tt arXiv:1301.6639 [hep-th]}}.

\bibitem{Ginsparg:1988ui}
P.~H. Ginsparg, ``{APPLIED CONFORMAL FIELD THEORY},'' in {\em {Les Houches
  Summer School in Theoretical Physics: Fields, Strings, Critical Phenomena Les
  Houches, France, June 28-August 5, 1988}}, pp.~1--168.
\newblock 1988.
\newblock \href{http://arxiv.org/abs/hep-th/9108028}{{\tt arXiv:hep-th/9108028
  [hep-th]}}.
\newblock
\url{https://inspirehep.net/record/265020/files/arXiv:hep-th_9108028.pdf}.
\newblock

\bibitem{Janik:2001hb}
R.~A. Janik, ``{Exceptional boundary states at c=1},''
  \href{http://dx.doi.org/10.1016/S0550-3213(01)00486-2}{{\em Nucl. Phys.} {\bf
  B618} (2001)  675--688},
\href{http://arxiv.org/abs/hep-th/0109021}{{\tt arXiv:hep-th/0109021
  [hep-th]}}.

\bibitem{Ginsparg:1987eb}
P.~H. Ginsparg, ``{Curiosities at c = 1},''
\href{http://dx.doi.org/10.1016/0550-3213(88)90249-0}{{\em Nucl. Phys.} {\bf
  B295} (1988)  153--170}.

\bibitem{Dixon:1988ac} 
  L.~J.~Dixon, P.~H.~Ginsparg and J.~A.~Harvey,
  Nucl.\ Phys.\ B {\bf 306}, 470 (1988).
  doi:10.1016/0550-3213(88)90011-9
  
\bibitem{Brunner:2010xm}
I.~Brunner and D.~Roggenkamp, ``{Attractor Flows from Defect Lines},''
  \href{http://dx.doi.org/10.1088/1751-8113/44/7/075402}{{\em J. Phys.} {\bf
  A44} (2011)  075402},
\href{http://arxiv.org/abs/1002.2614}{{\tt arXiv:1002.2614 [hep-th]}}.

\bibitem{Gukov:2015qea}
S.~Gukov, ``{Counting RG flows},''
  \href{http://dx.doi.org/10.1007/JHEP01(2016)020}{{\em JHEP} {\bf 01} (2016)
  020},
\href{http://arxiv.org/abs/1503.01474}{{\tt arXiv:1503.01474 [hep-th]}}.

\end{thebibliography}

\providecommand{\href}[2]{#2}\begingroup\raggedright\endgroup

\end{document}